\documentclass[pdflatex]{sn-jnl-arxiv}

\usepackage[utf8]{inputenc}
\usepackage[T1]{fontenc}
\usepackage{hyperref}
\usepackage{url}
\usepackage{booktabs}
\usepackage{microtype}
\usepackage{graphicx}
\usepackage{doi}
\usepackage{lmodern}
\usepackage{caption}
\usepackage{subcaption}

\usepackage{mathtools}
\usepackage{amsmath}
\usepackage{amsfonts}
\usepackage{amsthm}
\usepackage{amssymb}
\usepackage{centernot}
\usepackage{nicefrac}

\usepackage{multirow}
\usepackage{mathrsfs}
\usepackage[title]{appendix}
\usepackage{xcolor}
\usepackage{textcomp}
\usepackage{manyfoot}
\usepackage{listings}

\usepackage[capitalize]{cleveref}

\usepackage[numbers,sort&compress,square,comma]{natbib}

\title{Optimal Hospital Capacity Management During Demand Surges}

\author[1]{\fnm{Felix} \sur{Parker}}\email{fparker9@jhu.edu}
\author[1]{\fnm{Fardin} \sur{Ganjkhanloo}}\email{fganjkh1@jhu.edu}
\author[2,3]{\fnm{Diego A.} \sur{Martínez}}\email{dmart101@jh.edu}
\author[1]{\fnm{Kimia} \sur{Ghobadi}}\email{kimia@jhu.edu}

\affil[1]{
    \orgdiv{The Center for Systems Science and Engineering, Department of Civil and Systems Engineering, The Malone Center for Engineering in Healthcare},
    \orgname{Johns Hopkins University},
    \orgaddress{\city{Baltimore}, \state{Maryland}, \country{USA}}
}
\affil[2]{
    \orgdiv{School of Industrial Engineering},
    \orgname{Pontificia Universidad Católica de Valparaíso},
    \orgaddress{\city{Valparaíso}, \country{Chile}}
}
\affil[3]{
    \orgdiv{Department of Emergency Medicine},
    \orgname{Johns Hopkins University School of Medicine},
    \orgaddress{\city{Baltimore}, \state{Maryland}, \country{USA}}
}

\hypersetup{
	pdftitle={Optimal Hospital Capacity Management During Demand Surges},
	pdfsubject={},
	pdfauthor={Felix~Parker, Fardin~Ganjkhanloo, Diego A.~Martínez, Kimia~Ghobadi},
	pdfkeywords={hospital capacity management, hospital surge, healthcare logistics, COVID-19, operations research},
}

\newcommand{\T}{\mathcal{T}}
\newcommand{\C}{\mathcal{C}}
\renewcommand{\H}{\mathcal{H}}
\newcommand{\set}[1]{\{#1\}}

\begin{document}


\abstract{
Effective hospital capacity management is critical for enhancing patient care quality, operational efficiency, and healthcare system resilience, notably during demand spikes like those seen in the COVID-19 pandemic. However, devising optimal capacity strategies is complicated by fluctuating demand, conflicting objectives, and multifaceted practical constraints.
This study presents a data-driven framework to optimize capacity management decisions within hospital systems during surge events. Two key decisions are optimized over a tactical planning horizon: allocating dedicated capacity to surge patients and transferring incoming patients between emergency departments (EDs) of hospitals to better distribute demand. The optimization models are formulated as robust mixed-integer linear programs, enabling efficient computation of optimal decisions that are robust against demand uncertainty. The models incorporate practical constraints and costs, including setup times and costs for adding surge capacity, restrictions on ED patient transfers, and relative costs of different decisions that reflect impacts on care quality and operational efficiency.
The methodology is evaluated retrospectively in a hospital system during the height of the COVID-19 pandemic to demonstrate the potential impact of the recommended decisions. The results show that optimally allocating beds and transferring just 32 patients over a 63 day period around the peak, about one transfer every two days, could have reduced the need for surge capacity in the hospital system by nearly 90\%.
Overall, this work introduces a practical tool to transform capacity management decision-making, enabling proactive planning and the use of data-driven recommendations to improve outcomes.
}

\keywords{Hospital capacity management, Hospital surge, Healthcare logistics, COVID-19, Operations research}

\maketitle


\section{Introduction}
\label{sec:intro}

Effective management of hospital capacity is essential for improving the quality of care, controlling costs, and ensuring the resiliency of the healthcare system. High hospital capacity utilization rates can cause delays in patient care \citep{ofoma2020,forster2003a}, worsen outcomes for both patients and staff \citep{afugglas2020,morley2018,bosque-mercader2023,eriksson2017a,bernstein2009}, and even increase patient mortality risk \citep{ofoma2020,afugglas2020,bosque-mercader2023,eriksson2017a,fergusson2020}. However, there are significant financial pressures on hospitals to operate with high occupancy, and at some hospitals, this is often unavoidable due to increasingly frequent demand surges and the declining inpatient capacity of the healthcare system \citep{kelen2021, hospitalbasedemergencycareatthebreakingpoint2007, americanhospitalassociation2018}.
These competing care quality, operational, and financial objectives, coupled with the immense complexity of modern healthcare, many practical constraints, and constantly changing demands make effective capacity management extremely challenging and high stakes.
Given these challenges, quantitative methods that capture the full complexity of the problem and produce relevant information, insights, and recommendations can provide great value by helping hospital administrators improve their planning process and decisions.

Capacity management ultimately comes down to efficiently matching capacity and resources with patient demand. Broadly speaking, hospitals can increase the effective capacity of departments or services that need it, while limiting inflow or increasing outflow to reduce demand.
Increasing capacity by adding or opening more physical beds as needed is often not a viable solution due to cost, availability of staff and other resources, and regulatory constraints. Therefore, hospitals must decide how to distribute limited capacity between services and patient populations by reallocating beds, moving staff or modifying their schedules, and many other measures. They may also take steps to control demand, such as emergency department triage, transferring patients, canceling elective procedures, or implementing reverse triage to increase discharge rates.
Each of these interventions has complex effects on patient health outcomes, staff workload, and hospital operations. Determining the best decisions requires balancing the competing demands of quality and cost while taking into account the uncertain nature of healthcare demand.
While expert knowledge can be very valuable in understanding the impacts of these interventions, data-driven modeling can estimate them quantitatively, enabling objective, proactive planning and optimization of decisions.

Effective capacity management is particularly consequential in preparing for and responding to demand surges.
During surges, demand is highly elevated compared to usual levels, straining capacity. Furthermore, events that cause surges, such as pandemics or epidemics, natural disasters, and mass casualty events often lead to an influx of patients with similar characteristics and needs, putting specific departments or services under particularly acute stress and potentially pushing them to their breaking point.
Therefore, swift and effectual interventions with careful planning are of critical importance to ensure there are no major negative consequences on patient or staff health outcomes and hospital operations.
Capacity management for surge preparedness and response has become increasingly relevant in recent years, as many hospitals, particularly large academic medical centers, have faced a rising number of surge periods. This trend is the result of higher baseline occupancy levels coupled with major infectious disease outbreaks, natural disasters exacerbated by climate change, and other emergencies, without significant capacity increases \citep{americanhospitalassociation2018, delia2008}. In particular, the COVID-19 pandemic caused massive widespread hospitalization surges that pushed many hospitals to their limits, which highlighted the importance of capacity management during surges on an unprecedented scale.

The aim of this study is to produce practical tools for informing capacity management decisions during surge periods.
Recognizing that hospitals are complex systems and the decisions involved in capacity management are high-stakes, we do not attempt to fully automate the decision-making process. Instead, we aim to transform the way that humans make decisions by grounding them in data, projections, and optimized recommendations.
As there are too many capacity management strategies to capture in one model, we focus on specific strategies that are used in practice, are particularly effective during surge events, complement each other, and would benefit from proactive planning. We identified multiple decision types meeting these criteria and selected two of them.
At a high level, these strategies are: (1) allocating dedicated capacity for a surging patient population, and (2) transferring or diverting incoming patients between hospitals.
Allocating dedicated capacity for a surging patient population redistributes capacity within a hospital by setting aside dedicated capacity for a patient population that is causing a surge. Surge patient populations often have high priority or need specific resources that are shared or require setup; therefore, assigning them to dedicated capacity streamlines logistics and centralizes care, leading to better outcomes for all stakeholders. For these reasons, this strategy was used widely during the COVID-19 pandemic as COVID-19 patients required stringent infection control measures and specialized care.
Transferring patients boarding in the Emergency Department (ED) or diverting incoming patients from a hospital that is (or will soon be) out of capacity to hospitals that have spare capacity helps to balance demand within a hospital system, and ensures that hospitals do not run out of capacity \citep{painvin2021, pett2020}. It uses hospital system capacity cooperatively to reduce or even eliminate the need for dedicated surge capacity. Diversions and transfers were also used frequently throughout the COVID-19 pandemic to manage capacity during periods of peak stress \citep{hinson2022, garfinkel2021}.

In this study, we develop a data-driven optimization model to determine the optimal decisions that maximize the effectiveness of these capacity management strategies during surges.
The model uses demand predictions and capacity data to proactively plan decisions on what capacity should be dedicated to surge patients, and how many patients to transfer ahead of time, over any time frame or temporal resolution.
It captures a range of operational constraints, costs, and other real-world considerations involved in deploying these capacity management strategies so that the recommendations are practical and can be implemented without major adjustments.
We also make the model robust against the uncertainty in predicted demand so that it is reliable and trustworthy, even during volatile surge periods.
All of the practical considerations and the level of robustness are carefully parameterized so that decision-makers can have fine-grained control over the solutions, ensuring that the recommendations meet their needs while maintaining mathematical optimality.
Additionally, we develop a flexible method for forecasting hospital demand in certain surge settings that can supply the necessary predictions for proactive use of the decision optimization model in practice.

In \cref{sec:relatedwork}, we discuss other studies that investigate optimizing hospital capacity management. \Cref{sec:methods} introduces our methodology for predicting demand and optimizing dedicated capacity allocation and patient transfers during demand surges. We then apply our models to a case study using real-world data from a hospital system during the COVID-19 pandemic to demonstrate its potential benefits in \cref{sec:results}. Finally, we discuss the implications of this work and its limitations in \cref{sec:discussion}, and briefly summarize our conclusions in \cref{sec:conclusion}.

\section{Related Works}
\label{sec:relatedwork}

Our work centers on quantitative planning of hospital capacity management decisions, which has been explored extensively in the existing literature. Hospitals are highly complex systems with uncertain, time-varying demand and resources, and have many ways of managing capacity and demand, so there are many subproblems and approaches that have been investigated for the general problem of capacity management.
In this section, we discuss some of the existing methods for planning or optimizing capacity management decisions during both normal (\cref{sec:relatedwork:cm-nonsurge}) and surge periods (\cref{sec:relatedwork:cm-surge}), with a focus on methods that optimize capacity allocation or patient transfers. We also discuss methodologies for forecasting hospital demands (\cref{sec:relatedwork:forecast}).  

\subsection{Hospital Capacity Management During Non-Surge Periods}
\label{sec:relatedwork:cm-nonsurge}

Even during periods with typical demand, effectively managing hospital capacity and demand is essential to ensure hospitals use their capacity efficiently, which reduces costs and means patients receive timely and appropriate care. Many existing studies have looked at optimizing or improving some aspects of capacity management, including how to allocate beds between departments or services \citep{bekker2017, ayvaz2010a} or how to smooth demand \citep{helm2014, meng2015, chan2012, coffey2019, shi2021}.

Hospitals periodically redistribute capacity between departments or services to adapt to trends in demand or to improve efficiency. Several studies have investigated how to make decisions that best meet various objectives when reallocating capacity.
\citet{bekker2017} investigate four strategies for allocating beds between different patient populations with the aim of improving flexibility to adapt to normal demand fluctuations.
\citet{ayvaz2010a} develop a dynamic programming model for allocating surgical capacity between patients seeking elective care and patients seeking emergency care. While broadly similar to allocation of dedicated capacity, their method does not take into account patient length of stay or other practical considerations, which are of critical importance in implementation.
\citet{aslani2021} consider tactical capacity planning in an outpatient setting, dividing capacity between first and return-visit patients using a robust optimization model.

There are also several studies that aim to improve the efficiency of hospital operations by smoothing demand.
\citet{helm2014} develop an optimization model for planning elective admissions with the aim of stabilizing a hospital census.
\citet{ridge1998} explore the relationship between capacity utilization and patient transfers using a simulation model, addressing the complexities of capacity allocation in a hospital setting.
\citet{shi2021} focus on discharges rather than admissions, and develop a decision-support framework for optimizing when patients should be discharged to reduce hospital congestion without harming patient outcomes.
Finally, \citet{nezamoddini2016} propose a method for optimally transferring patients between hospitals with the aim of reducing wait times in a non-surge setting. They develop a granular model that makes hourly decisions on admissions, discharges, and transfers during one 8-hour shift.

For further discussion of the literature, see \citet{humphreys2022}, which reviews many additional studies that relate to optimizing hospital capacity management.
While these studies are relevant to our work, they do not consider the short-term interventions that are essential for tactical capacity management planning during surges or the specific challenges posed by surge periods.

\subsection{Hospital Capacity Management During Surge Periods}
\label{sec:relatedwork:cm-surge}

More directly relevant to our work is the literature on how to manage the capacity of a hospital during surges. There are many interventions that hospitals can make to respond to a surge, each of which can interact with others to increase the strained capacity or reduce demand. Numerous papers have looked at a subset of these interventions and built models to simulate or optimize decision-making for these interventions. This area of study gained a lot of attention during the COVID-19 pandemic, which highlighted the need for effective surge response.
\citet{melman2021} explored this through a simulation model that examined the impact of canceling elective procedures and dividing operating room capacity between COVID-19 and elective surgeries.
Additionally, \citet{wood2020} modeled capacity increases, length of stay decreases, and admission limits as strategies to reduce capacity-related mortality risks, underscoring the need for robust demand management models.
\citet{toerper2018} developed an online dashboard for decision-makers to simulate the effects of a surge and response strategies. Their tool allows for exploring the effects of opening unlicensed beds, canceling elective procedures, and implementing reverse triage.

\citet{bai2014} builds a model to optimize patient distribution between hospitals during a surge period. However, instead of optimizing transfers, they optimize incentives to encourage patients to align their choice of hospital with the optimal distribution of patients within the system. Such an approach can greatly improve the distribution of load between hospitals during surges without requiring difficult and costly transfers, but additional work is needed on incentive design and deployment to make it practical.
While there are many studies that investigate optimizing hospital capacity management during surge periods, we are not aware of any that optimize the allocation of dedicated surge capacity. As this is a common tactic that is a major component of surge response in hospitals, we feel that it represents a significant gap in the literature, which we aim to fill in this study.

\subsubsection*{Optimizing Patient Transfers}
Our work is particularly related to the existing literature on modeling inter-hospital patient transfers, which has received attention as an important capacity management tool to study, particularly since the beginning of the COVID-19 pandemic.
\citet{lacasa2020} develop a model for load-sharing of Intensive Care Unit (ICU) patients in a hospital system using random search optimization and apply it to health systems in the United Kingdom and Spain at a regional level for the COVID-19 pandemic.

\citet{acuna2022} introduce bargaining optimization frameworks for diminishing healthcare waiting times by enabling patient trades and private hospital integration in two-tier healthcare systems. They employ game-theoretic models, calibrated with data from Chile, to assess the efficacy of these strategies in fostering public-private collaborations and facilitating patient exchanges among hospitals. The analysis reveals substantial potential for reducing waiting lists through these innovative approaches.
\citet{marquinez2021} introduce a method for making transfer decisions in the Chilean public hospital system. They model the problem of deciding how many patients to admit, transfer to another public hospital, or divert to a private clinic at each public hospital as a Markov decision process and solve it approximately using a dynamic programming algorithm.
\citet{sun2014} develop integer programming models for determining an optimal allocation of patients and resources to hospitals in an influenza outbreak that minimizes the total distance that patients have to travel.
Finally, and closest to our work, \citet{parker2020} develop a series of linear programming models to optimize the redistribution of resources and demand between nodes in a network, and apply these models to optimizing patient transfers during the COVID-19 pandemic. Our model of patient transfers is based on their formulation of the problem, but we extend their method in a number of critical ways, including incorporating it with our model of capacity allocation and better handling of uncertainty from forecasts.

Although these existing methods seek to optimize patient transfers, they overlook capacity optimization, a crucial aspect considering the practical challenges and limits of implementing transfers. Moreover, these approaches often miss key practical constraints and costs linked to patient transfers, undermining the feasibility of their recommendations.
Additionally, there is a notable absence of prospective validation and the necessary data infrastructure for real-time and prospective application in hospital decision-making. Our work addresses these shortcomings by proposing a practical, forward-looking, data-driven system for optimizing both patient transfers and capacity allocation during demand surges.

\subsection{Hospital Demand Forecasting}
\label{sec:relatedwork:forecast}

A crucial aspect of capacity management, particularly during demand surges, is the ability to accurately forecast demand. Demand predictions can directly inform many capacity management decisions including how many nurses to schedule and how much total capacity is required.
Hospital demand forecasting has received much attention, particularly in the wake of the COVID-19 pandemic \citep{johnson2023, baas2021, deschepper2021, klein2023, kociurzynski2023, taghia2022, vollmer2021, weissman2020, yaesoubi2023, yang2021a}.
\citet{johnson2023} demonstrate the importance of predictions in their study on predicting unit-level patient demand during the COVID-19 pandemic. Their work shows that using demand forecasting could lead to better capacity management decisions. Implemented at two hospitals, their study also included a user-friendly dashboard to assist in decision-making, highlighting the practical application of such models.

Many studies have developed and evaluated quantitative forecasting methods to predict hospitalizations in a variety of settings.
\citet{jones2008} explore and evaluate various statistical forecasting methods to predict daily ED patient volumes at three hospitals. They find that regression-based models incorporating calendar variables, site-specific special days, and residual autocorrelation provide the most accurate approach.
\citet{weissman2020} present a modeling tool called the COVID-19 Hospital Impact Model to estimate the timing and intensity of COVID-19 surges to inform hospital capacity planning. The model provides best- and worst-case scenarios to identify when hospital capacity could become saturated.
\citet{zhu2017} compare time series models for forecasting daily hospital discharges. They find seasonal regression combined with ARIMA performs well, and these accurate discharge forecasts can inform admission scheduling and capacity planning decisions.
\citet{ordu2021a} develop a hybrid forecasting-simulation-optimization model to estimate required bed capacity and staffing needs across all hospital units and services. Their model highlights the benefits of linking forecasting, simulation, and optimization.
\citet{baas2021} present a real-time forecasting model to predict COVID-19 ward and ICU occupancy based on predicted admissions, length of stay, and patient transfers. Tested during the COVID-19 pandemic in the Netherlands, their model provided highly accurate occupancy forecasts to inform hospital capacity management.
The literature highlights the challenge and relevancy of developing quantitative methods to produce accurate predictions of hospital demand and inform capacity planning, especially during surge periods. Further development and application of such models can greatly assist in hospital capacity management.

\subsection{Contributions}
Our work makes several contributions to the literature on hospital capacity management.
First, we develop a model for optimizing dedicated capacity allocation decisions during demand surges. Allocating dedicated capacity for surge patients is one of the primary strategies that hospitals use to manage large surges, however, there is little existing work on how to optimize the corresponding decisions.
Second, we integrate dedicated capacity allocation with a patient transfer optimization model to take advantage of the complimentary nature of these approaches, with capacity allocation increasing the necessary supply and transfers reducing the peak demand. Existing work on optimizing transfers is not practical enough for real-world use and does not consider the interactions between these strategies.
Third, we establish a method for facility-level probabilistic demand forecasting that achieves strong performance, enabling prospective decision optimization.
To ensure decisions are reliable and trustworthy when using uncertain demand predictions, we introduce a robust extension of the optimization model. We also incorporate a variety of operational considerations into the model, with a focus on real-world implementation of capacity management strategies, so that it can provide real practical value to decision-makers.
Finally, we validate the viability and benefits of this approach to capacity management using a real-world hospital data from a COVID-19 surge period.

\section{Methodology}
\label{sec:methods}

\begin{figure*}[ht]
	\centering
	\includegraphics[width=0.85\textwidth]{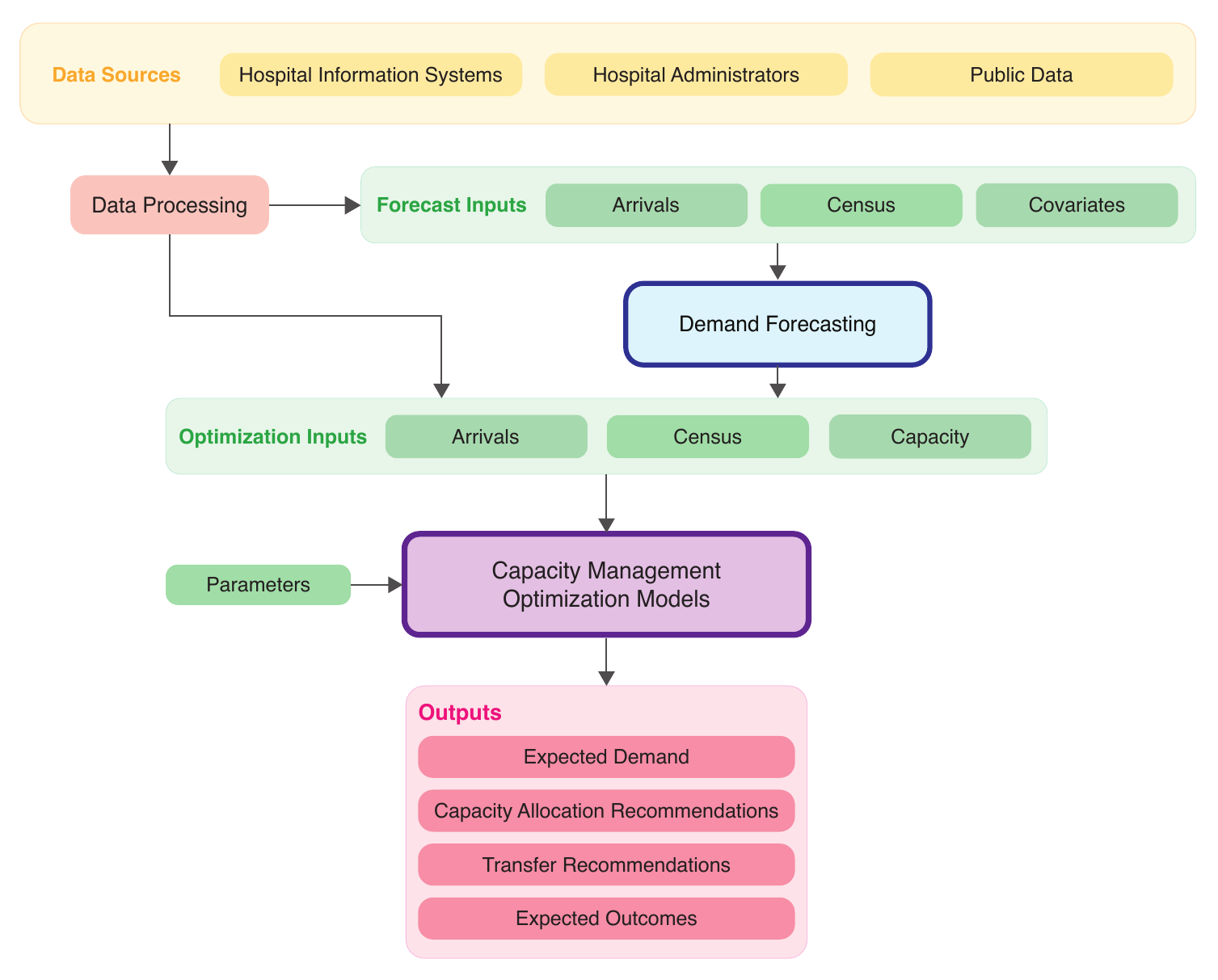}
	\caption{Overview of our framework for optimizing capacity management decisions. The primary component of the framework and the focus of this work is the set of capacity management models (purple). It is data-driven and prospective, so it requires data from the hospital system (green) as well as forecasts of future demand (blue). This data can come from a variety of sources (yellow).}
	\label{fig:method-overview}
\end{figure*}

This study develops a practical, prospective, and data-driven framework for optimizing certain hospital capacity management decisions during surge events. It can provide decision-makers with estimates of future surge demand, quantify the impact of surges on hospital operations, and recommend optimal plans to manage capacity that are robust and tailored to the practical needs and considerations of the hospital system. It does so using the components shown in \cref{fig:method-overview}: data collection and processing (described in \cref{sec:methods:data}), forecasting (described in \cref{sec:methods:forecast}), and decision optimization (the main focus of this paper; described in \crefrange{sec:methods:capacity}{sec:methods:uncertainty}). Each of these components provides information that is useful for capacity management, but the primary contributions and focus of this work are the decision optimization models.

The first capacity management strategy included in our model is increasing the available resources for a surging patient population by allocating additional dedicated capacity to them. The specific decisions we optimize are which groups of beds to allocate as dedicated capacity for the surge patients during each time period in the planning horizon.
Dedicated capacity for surging patient populations can be allocated at different levels of granularity depending on hospital infrastructure, clinical needs of the surge patient population, and available data. These levels can include individual beds, units, sections of units, or floors. To simplify the terminology, throughout the remainder of this paper, we will interchangeably use ``units of capacity'', or just ``units'', to refer to arbitrary groups of beds that will be allocated and converted together, but do not necessarily correspond to hospital units. These units are identified by hospitals and departments based on logistical considerations.

The second capacity management strategy we include in our model is redistributing demand by transferring or diverting patients between hospitals. The model determines the optimal number of patients to transfer from each hospital to each of the other hospitals in the system during each time step.
The primary set of patients we consider for transfer are those who have arrived at an ED, are in the surge patient population, and require inpatient care, but have not yet been assigned an inpatient bed.
This group consists primarily of ED boarders, patients who have a pending admission order but are waiting in an ED bed until they are assigned an inpatient bed. During surges, as inpatient capacity fills up, the number of ED boarders can rise rapidly. Extended ED boarding can be harmful to patients and adds to stress on the ED; therefore, transferring these patients can benefit the patients and the hospitals.
We also consider patients who are on their way to a particular hospital for diversion. Diversions are commonly done through alerts to regional emergency medical service (EMS) systems telling them that a hospital is full.
We do not distinguish between transfers and diversions in our model as they both serve to shift demand that would be served by one hospital to another hospital. Both are also used in practice during demand surges.
We do not consider transfer of inpatients because these have less impact on capacity utilization and require strong clinical motivations.

It is essential that both capacity management strategies are planned ahead of time so that hospitals have the time they require to implement any recommended decisions, including converting capacity, adjusting staffing, and planning patient transport. This requires advance predictions of demand so that our optimization models can determine the impacts of decisions and optimize them.
Early warning about the timing and severity of surges is also extremely helpful to hospitals as it enables them to make informed decisions beyond these strategies and generally ensure they are well prepared.
We employ a data-driven approach to hospital-level demand forecasting with uncertainty estimates using machine learning. Our methodology is described further in \cref{sec:methods:forecast}.

Our methodology is most suited for medium-term tactical planning for a single surge, spanning roughly two to eight weeks. Over this time frame, demand forecasts can be sufficiently accurate, hospitals have time to prepare for and implement decisions, and low-level implementation details are not the primary focus. Many types of surges including infectious disease waves and natural disasters fall within this window.
For tactical planning, optimizing decisions at daily granularity is most useful as it strikes a good balance between detail and flexibility, aiding in practical implementation, and does not require more granular forecasts which are less accurate.
However, our model can be used with longer or shorter time steps as needed. Similarly, it can be useful at both longer and shorter time horizons for operational or strategic planning, although it is not built around those use cases.

Our model optimizes decisions for a system of hospitals in which transfers can be made in a cooperative manner. Most commonly, such a system would consist of hospitals in the same geographic region that are owned by the same health system, and therefore do not compete as directly as others. However, during particularly large surges, such as waves of the COVID-19 pandemic, there can be cooperation between a wider range of hospitals.
Capacity allocation decisions can be made by hospitals independently, and this is supported by our model, but once transfers are allowed the optimal allocation decisions will start to depend on the number of transfers, and vice versa, so these decisions cannot be decoupled and the whole system must be included.

\subsection{Data Collection}
\label{sec:methods:data}

A key feature of our methodology is that it is data-driven, so all predictions and recommendations are grounded in real-time and historical data from a hospital system. Modeling a hospital system in a data-driven manner is critical to ensure that the selected decisions are feasible, effective, and optimal, particularly during demand surges.
Specifically, the data that our optimization models require consists of: (1) the number of patient arrivals from the target population at each hospital during each time step, (2) the resulting census, and (3) the capacity of each bed-group that can be allocated to the target patient population.
In a practical setting, these inputs must be determined prospectively, with a forecast of admissions and census, and a plan for surge expansion of capacity.
However, to enable data-driven predictions, retrospective data for these inputs is necessary. We discuss our approach to forecasting demand further in \cref{sec:methods:forecast}, but its primary inputs are historical patient admissions and census for our target population.
Furthermore, for retrospective evaluation of our methodology in \cref{sec:results}, we also utilize historical demand data.

Retrospective data on admissions and census can be collected from Hospital Information Systems (HIS) or Electronic Health Records (EHR). The details of these systems vary by hospital, but there are certain common considerations.
Records must be filtered to only include patients in the target patient population for dedicated capacity allocation and transfers. Records must also be aggregated to the desired temporal resolution, typically daily. The sum of admissions in a time step should be used, while the peak census, or the census at a time of day when hospitals are typically busiest, is more appropriate because hospitals must have enough capacity for the peak census.
Our models use aggregate admissions and census data rather than relying on individual patient data, and therefore do not require sensitive information, and can be solved quickly even for large patient volumes.

The capacity of each unit or group of beds that can be allocated to the target patient population must be known in order to make capacity allocation decisions. The number of staffed beds in each group can be collected from HIS or other hospital records.
However, some decisions need to be made on what level of granularity is appropriate for allocating capacity (individual beds, rooms, units, etc.), and which groups of beds should be considered as possible sites for dedicated surge capacity. These are complex decisions that must be made ahead of time by hospital administrators as they depend on the specific needs of the surge patient population and each hospital's infrastructure. While these decisions are not directly addressed by our model because they are too hospital-specific, hospital administrators can experiment with different inputs to our model to test their strategies, observe how the recommendations change, and validate their decisions.

One of the additional necessary inputs to our transfer optimization model is an estimate of the distribution of patient length of stay (LOS), as it allows us to estimate the census of a hospital based on past admissions and transfers. Patient LOS can have significant variability, and is difficult to predict for individual patients, but we typically can accurately estimate the marginal probability distribution over LOS.
If the LOS for patients previously discharged is available, the distribution can be estimated using maximum likelihood estimation.
Otherwise, the LOS distribution can be approximated using aggregate admissions and census data.
\Crefrange{eq:exp:census}{eq:exp:discharges} describe the relationship between census, admissions, and LOS. If census and admissions are known, this model can be used to estimate the LOS distribution by minimizing a difference measure between the predicted census and true census with respect to the distribution parameters.

\subsection{Capacity Allocation}
\label{sec:methods:capacity}

In this section, we detail our model for optimizing the allocation of dedicated capacity to a surge patient population. Specifically, we use demand data or predictions to optimize the selection of units or groups of beds to reserve at each point in time.
We assume patients from this population have similar needs in terms of resources, level of care, infection controls, or other characteristics that distinguish them from other patients, which means dedicated capacity is logistically beneficial. These assumptions are tailored to epidemic or pandemic conditions, and inspired by learnings from the COVID-19 pandemic, but are applicable to many surge settings.
On the surface, deciding which units to allocate to a group of patients may seem straightforward, but due to highly uncertain demand, many operational considerations, complex interactions with other decisions, and the atypical stresses caused by surges, it is a challenging problem that benefits significantly from quantitative, data-driven planning and optimization.

We aim to model the complex reality of the problem. We incorporate realistic operational considerations in our models as constraints and objectives term, including:
\begin{enumerate}
	\item Capacity shortage can significantly impact patient care and should be avoided as much as possible.
	\item Cost (monetary and otherwise) should be minimized, but not at the expense of patient care.
	\item Allocating capacity to a specific patient population takes away potential capacity from other patients.
	\item Allocating capacity involves converting (and re-converting) units, which requires downtime and carries cost that may depend on the unit and vary over time (see \crefrange{eq:cons:allocated:a}{eq:cons:conversion:c}). 
	\item Reaching 100\% occupancy is impractical in hospitals.
	\item Patient demand forecasts are inherently uncertain, and good decisions must be robust against this uncertainty.
	\item Capacity allocation is interconnected with other capacity management decisions, which should be taken into account.
\end{enumerate}

\begin{table*}
	\centering
	\begin{tabular}{p{2cm}p{4cm}p{9cm}}
		\toprule
		Notation			& Name		& Description		\\
		\midrule
		\multicolumn{3}{c}{\textit{Sets}} \\
		\midrule
		\( \T \)		& Time steps		& Set of all time steps (days)		\\
		\( \H \)		& Hospitals		& Set of hospitals		\\
		\( \C \)		& Bed groups		& Set of bed groups or units of capacity		\\
		\midrule
		\multicolumn{3}{c}{\textit{Data}} \\
		\midrule
		\( o_{h,t} \)		& Census		& Census of hospital $h$ at time $t$		\\
		\( b_{h,k} \)		& Capacity per unit		& Number of staffed beds at hospital $h$ in unit $k$		\\
		\( i_{h,t} \)		& Incoming demand		& Number of patients arriving at hospital $h$ during time period $t$		\\
		\( \mathcal{L}_h \)		& Length of stay		& Distribution over length of stay for patients at hospital $h$		\\
		\midrule
		\multicolumn{3}{c}{\textit{Variables}} \\
		\midrule
		\( c_{h,t} \)		& Allocated capacity		& Operational capacity allocated at hospital $h$ during time period $t$		\\
		\( u_{h,t,k} \)		& Unit usage		& Binary variable indicating whether unit $k$ at hospital $h$ is in use during time period $t$		\\
		\( s_{h,g,t} \)		& Patient transfers		& The number of patients to transfer/divert from hospital $h$ to hospital $g$ during time period $t$		\\
		\midrule
		\multicolumn{3}{c}{\textit{Parameters}} \\
		\midrule
		\( w_1, w_2, w_3 \)		& Objective function weights		& Costs or weights for the optimization objective function		\\
		\( S \)		& Transfer limits		& Upper bounds on the number of patients that can be transferred		\\
		\( \Gamma_1, \Gamma_2 \)		& Uncertainty set parameters		& Control the deviation of uncertain scenarios from the nominal predictions	\\
		\bottomrule \\
	\end{tabular}
	\caption{Mathematical notation used in the optimization model developed in \cref{sec:methods:capacity}. Only the most relevant notation is included for brevity. Additional notation is introduced as needed.}
	\label{tab:notation}
\end{table*}

In order to solve this decision optimization problem in full detail we model it as a mixed-integer linear program (MILP). We have found that this structure allows us to faithfully model the problem and solve for globally optimal decisions efficiently.
Table \ref{tab:notation} defines the notation used in formulating the model.
Our primary decision variable, binary variable \( u_{h,t,k} \), indicates whether a unit $k$ at hospital $h$ is allocated and available as dedicated capacity during time period $t$ (which we refer to as day $t$, but can represent any period of time). Each unit has a corresponding capacity \( b_{h,k} \), which depends on physical beds, staff, and other resources. This can be used to determine the total allocated capacity at a hospital on a day, \( c_{h,t} \)  as shown in \cref{eq:exp:capacity}.
Constraint \eqref{eq:cons:no-shortage} ensures that the total available dedicated capacity is not smaller than the census value, \( o_{h,t} \), at any time $t$, and hence, there is no shortage of care.
\begin{subequations}
\begin{align}
	&c_{h,t} = \sum_{k \in \mathcal{C}} u_{h,t,k} b_{h,k}		&\forall h \in \H, t \in \T	\label{eq:exp:capacity} \\
	&o_{h,t} \leq c_{h,t}			&\forall h \in \H, t \in \T	\label{eq:cons:no-shortage} \\
	&u_{h,t,k} \in \set{0,1}			&\forall h \in \H, t \in \T, k \in \C
\end{align}
\end{subequations}

Since in practice it takes time to convert capacity, and there are costs associated with doing so, we introduce variables \( \hat{u}_{h,t,k} \) which represent when the corresponding unit is allocated, not just usable as shown in constraints \eqref{eq:cons:allocated:a}--\eqref{eq:cons:allocated:d}. Similarly, we introduce variables \( \check{u}_{h,t,c} \) to track when a unit of capacity is converted back (constraints \eqref{eq:cons:conversion:a}--\eqref{eq:cons:conversion:c}. The variable \( \hat{u}_{h,t,k} \) is activated (has value 1) for $\delta^-_{h,k}$ days before and $\delta^+_{h,k}$ days after unit $c$ at hospital $h$ is available, where $\delta^-_{h,k}$ is the number of days it takes to set up the dedicated capacity, and $\delta^+_{h,k}$ is the number of days to convert back.
The variable \( \check{u}_{h,t,k} \) is activated if unit $k$ at hospital $h$ is allocated for day $t$ but not day $t-1$, indicating that it must be converted.
\begin{subequations}
\begin{align}
	u_{h,t,k} &\leq \hat{u}_{h,t-\delta^-_{h,k},k}		&\forall h \in \H, t \in \T, k \in \C	\label{eq:cons:allocated:a} \\
	u_{h,t,k} &\leq \hat{u}_{h,t+\delta^+_{h,k},k}		&\forall h \in \H, t \in \T, k \in \C	\label{eq:cons:allocated:b} \\
	u_{h,t,k} &\leq \hat{u}_{h,t,k}		&\forall h \in \H, t \in \T, k \in \C	\label{eq:cons:allocated:c} \\
	\hat{u}_{h,t,k} &\in \set{0,1}			&\forall h \in \H, t \in \T, k \in \C	\label{eq:cons:allocated:d}\\
\end{align}
\end{subequations}
\begin{subequations}
\begin{align}
	\check{u}_{h,t,k} &\geq u_{h,t,k} - u_{h,t-1,k}		&\forall h \in \H, t \in \T \setminus \set{1}, k \in \C	\label{eq:cons:conversion:a} \\
	\check{u}_{h,1,k} &= u_{h,1,k}		&\forall h \in \H, k \in \C	\label{eq:cons:conversion:b} \\
	\check{u}_{h,t,k} &\in \set{0,1}			&\forall h \in \H, t \in \T, k \in \C	\label{eq:cons:conversion:c}
\end{align}
\end{subequations}

We can construct an objective function (\cref{eq:obj}) that measures the total cost of allocating dedicated capacity. This includes the cost of reserving each bed, the cost of reserving each unit, and the cost of converting each unit (which may vary by hospital $h$ and unit $k$), denoted by $w_{1,h,k}$, $w_{2,h,k}$, and $w_{3,h,k}$ respectively. It is important to note that these do not necessarily represent financial costs -- there are complex costs to hospital operations, staff, and patient care that must be accommodated.
Determining monetary costs and patient harms is outside the scope of this work, but appropriate relative costs can be readily determined for most settings and iteratively refined as needed.
\begin{equation}
    \text{min} \sum_{t \in \T} \sum_{h \in \H} \sum_{k \in \C} \left( w_{1,h,t} c_{h,t} + w_{2,h,k} \hat{u}_{h,t,k} + w_{3,h,k} \check{u}_{h,t,k} \right)
	\label{eq:obj}
\end{equation}

Finally, we incorporate some optional practical constraints that can be used as needed. Instead of carefully estimating the relative costs of allocating different units of capacity, it may be simpler in practice to assign them a priority, and then allocate them only in priority order. Constraint \eqref{eq:cons:unit-order} ensures units are allocated in order, meaning that unit 3 can only be allocated if units 1 and 2 are as well.
Hospitals are also generally unable to reach 100\% occupancy, and doing so would not even be desirable due to the inefficiencies it would create. Constraints \eqref{eq:cons:util-rate-1} and \eqref{eq:cons:util-rate-2} enable users to set the maximum occupancy ($z$) in different ways.
\begin{subequations}
\begin{align}
    &u_{h,t,k} \leq u_{h,t,k-1}  &\forall h \in \H, t \in \T, k \in \C
	\label{eq:cons:unit-order}\\
    \
    &o_{h,t} \leq z c_{h,t}		&\forall h \in \H, t \in \T		\label{eq:cons:util-rate-1} \\
	&o_{h,t} \leq c_{h,t} - z^\prime		&\forall h \in \H, t \in \T		\label{eq:cons:util-rate-2}
\end{align}
\end{subequations}

There are many more hospital-specific considerations that can be easily added to the model as needed due to the flexibility of our approach.
We have also not addressed the uncertainties in demand or the interactions with other decision types, which we will discuss in the remainder of \cref{sec:methods}.

\subsection{Patient Transfer Optimization}
\label{sec:methods:transfers}

In this section, we extend the model to incorporate patient transfers. 
There is considerable existing literature on optimizing patient transfers, including on linear models that are compatible with our framework \citep{parker2020,sun2014}.
The formulation proposed by \citet{parker2020} optimizes transfer of demand between nodes in a network, with the aim of minimizing the resulting resource shortage. While we model capacity differently so that there is no shortage allowed, their formulation can be readily adapted to suit the scenario we consider for patient transfers. We therefore adapt their formulation and extend it.
In the remainder of this section we describe the adapted model.

The formulation of this model involves some mild assumptions, which include that we only consider transfer of arriving patients, and that these transfers are completed within a single time period. We discuss the assumptions made in greater detail in \cref{sec:discussion}.
It is important to note that the model only determines the optimal number of patients to transfer, not the specific individuals that should be transferred. Those decisions are based on clinical factors and must be made by clinicians.

We employ the patient transfer formulation as an extension of the capacity allocation model, rather than a separate model, so decisions are optimized jointly, with the same objective function.
We introduce a new decision variable, $s_{h,g,t} \in \mathbb{Z}_{\geq 0} \, \forall h \in \H, g \in \H, t \in \T$, to represent the number of incoming patients to transfer from hospital $h$ to hospital $g$ during day $t$. This variable is constrained so that the sum of outgoing transfers from a hospital on a given day is no more than the number of incoming patients, ensuring that no inpatients are considered for transfer, i.e., $\sum_{g \in \H} s_{h,g,t} \leq i_{h,t},~\forall h \in \H, t \in \T$. 
We augment the objective function (\cref{eq:obj}) to reflect the costs of transfer with additional terms $\sum_{h \in \H} \sum_{g \in \H} \sum_{t \in \T} w_{4,h,g} s_{h,g,t}$.

Allowing for transfers, the census of each hospital will then depend on the previous transfer decisions, so it cannot be predicted directly. Instead, we use a forecast of incoming patients to each hospital during each day, $i_{h,t}$, and estimate admissions, discharges, and census using this information along with the recommended transfers.
Admissions are computed from the number of patient arrivals and the net transfers \eqref{eq:exp:admissions}, and discharges are estimated based on previous admissions and patient LOS \eqref{eq:exp:discharges}. The census is then determined by the admissions and discharges on all previous days \eqref{eq:exp:census}.
In particular, the expected number of discharges in a given day, $t$, is the number of admissions on previous days, $t^\prime \leq t$, multiplied by the probability that a patient admitted on day $t^\prime$ was discharged on day $t$, summed over all previous days $t^\prime$.
When used in conjunction with the capacity allocation model, this estimate of patient census takes the place of the census data. This estimate can also be used in place of census data, even without transfers, in settings in which only arrival data or forecasts are available.
\begin{subequations}
\begin{align}
	o_{h,t} &= \sum_{t^\prime = 1}^{t} \left( a_{h,t^\prime} - d_{h,t^\prime} \right)		&\forall h \in \H, t \in \T		\label{eq:exp:census} \\
	a_{h,t} &= i_{h,t} + \sum_{g \in \H} \left( s_{g,h,t} - s_{h,g,t} \right)			&\forall h \in \H, t \in \T		\label{eq:exp:admissions} \\
	d_{h,t} &= \sum_{t^\prime = 1}^{t} \left( P(\mathcal{L}_h = t - t^\prime) a_{h,t^\prime} \right)		&\forall h \in \H, t \in \T	\label{eq:exp:discharges}
\end{align}
\end{subequations}

Existing systems within hospitals are not set up to transfer a significant number of patients consistently. Therefore, hospitals will generally want to limit the number of transfers recommended by the model. We add constraints on the number of transfers that reflect limits on the transfers that can be done in a day, or over the entire time span of the model (\cref{eq:cons:transfer-limits}).
These limits can be set by decision-makers to control the solution so that the number of transfers made is feasible in practice.
Additional operational considerations are discussed in \citet{parker2020}.
\begin{subequations}
\begin{align}
	s_{h,g,t} \leq S_{h,g}, \quad \sum_{g \in \H} s_{h,g,t} \leq S_{h}, \quad \sum_{h,g \in \H} \sum_{t \in \T} s_{h,g,t} \leq S	\qquad\quad	\forall h \in \H, g \in \H, t \in \T    \label{eq:cons:transfer-limits}
\end{align}
\end{subequations}

\subsection{Uncertainty Management}
\label{sec:methods:uncertainty}

Properly managing the uncertainty that is inherent in prospective planning is essential to ensure that the recommended decisions are valuable and do not result in harmful outcomes. If we optimize for the expected future demand, but the true demand surges much higher than expected, a hospital may be unprepared, which can have adverse effects on patients. To ensure that the recommended decisions account for uncertainty and eliminate the potential for harm, we turn to robust optimization, which guarantees that the robust optimal decision is feasible for any possibility in a selected uncertainty set \citep{bertsimas2011}.
In this section, we focus on uncertainties in future demand, specifically the patient census and the number of incoming patients at each hospital during each day in the future. However, we will also discuss uncertainty in the other inputs to the decision optimization models.

In order to provide a strong guarantee of solution feasibility without making the resulting decisions too conservative to be effectual, we develop an uncertainty set with a customizable `budget of uncertainty' \citep{bertsimas2004, bertsimas2011} that is tied to the specifics of the problem setting. It utilizes knowledge of how demand evolves over time and the past performance of the forecast to exclude extreme scenarios.
First, \cref{eq:exp:uncertainty-set-census} constrains the uncertain values to be between the lower and upper bounds dictated by the forecast. It then eliminates scenarios in which the forecast would have a Mean Absolute Percent Error (MAPE) of more than $\Gamma_1$, a budget of uncertainty. This is particularly useful when the historical performance of the forecast is known, as we can assume future performance will be similar, which provides information for setting $\Gamma_1$.
The uncertainty set then has a constraint on how much demand can change between consecutive days, normalized by how much it was expected to change. This eliminates unlikely scenarios with huge fluctuations or sudden spikes.
These assumptions about the performance of the forecast and behavior of demand allow decision-makers to improve the utility of the decisions without being too concerned about not being robust against scenarios that may actually occur.
\begin{equation}
	O_h = \left\{  o_h \in \mathbb{R}^{|\T|} \big\vert o_{h,t} \in [\bar{o}_{h,t} - \tilde{o}^-_{h,t}, \bar{o}_{h,t} + \tilde{o}^+_{h,t}], 
	\frac{\sum_{t \in \T} \left| o_{h,t} - \bar{o}_{h,t} \right|}{\sum_{t \in \T} \bar{o}_{h,t}} \leq \Gamma_1, 
		\left| \frac{o_{h,t} - o_{h,t+1}}{\bar{o}_{h,t} - \bar{o}_{h,t+1}} \right| \leq \Gamma_2, \quad\forall t \in \T
        \right\}
	\label{eq:exp:uncertainty-set-census}
\end{equation}

If transfers are optimized, the census forecast cannot be used directly, so $O_{h}$ must be replaced by $O^{\prime}_{h}$, which is computed from the uncertainty set for incoming demand $I_{h}$. The only additional change is that the constraint on outgoing transfers from a hospital $h$ during a time step $t$ uses $i_{ht}$ as an upper bound, so those values must be replaced by \( \tilde{i}_{ht} = \left( \min \set{ i_h : i_h \in I_h } \right)_t \).

Solving a robust optimization model requires ensuring each constraint remains feasible for each possible value in the uncertainty set. In general, this can be complex, but we structured the problem such that the model can be solved simply and efficiently.
The uncertain values, census $o_{ht}$ and incoming demand $i_{ht}$, only appear in constant terms of the model. Occupancy appears only in \cref{eq:cons:no-shortage,eq:cons:util-rate-1,eq:cons:util-rate-2}, where it serves as part of a lower bound for a function of variables. Therefore, to ensure these constraints are not violated for any possible values of $o_{ht}$, it suffices to ensure that the constraint holds for \( \tilde{o}_{ht} = \left( \max \set{ o_{h} : o_{h} \in O_{h} } \right)_t \) because \( o_{ht} \leq \tilde{o}_{ht} \, \forall \, o_{h} \in O_{h} \).
Finding $\tilde{o}_{ht}$ involves solving this extremely small and simple Linear Program (LP), so solving the robust optimization model reduces to first solving $|\H|$ small auxiliary LPs and then solving the original MIP with the auxiliary results plugged in for $o_{ht}$. The auxiliary LPs take negligible time to solve in practice, and even in the worst case can be solved in polynomial time.
\begin{subequations}
\begin{align}
	I_h &= \left\{ 
             i_h \in \mathbb{R}^{|\T|} :
             i_{h,t} \in [\bar{i}_{h,t} - \tilde{i}^-_{h,t}, \bar{i}_{h,t} + \tilde{i}^+_{h,t}], 
		 \frac{\sum_{t \in \T} \left| i_{h,t} - \bar{i}_{h,t} \right|}{\sum_{t \in \T} \bar{i}_{h,t}} \leq \Gamma_1,
		\left| \frac{i_{h,t} - i_{h,t+1}}{\bar{i}_{h,t} - \bar{i}_{h,t+1}} \right| \leq \Gamma_2, \quad\forall t \in \T
         \right\}
	\label{eq:exp:uncertainty-set-incoming}\\
    \
	O^\prime_h &= \left\{ \tilde{o}(i_h, s, \mathcal{L}_h) : i_h \in I_h \right\}
	\label{eq:exp:uncertainty-set-census-prime}
\end{align}
\end{subequations}

\citet{parker2020}, the basis for our patient transfer model, also employs a robust optimization model with a budget of uncertainty. However, their uncertainty set makes an entirely different set of assumptions about how demand can differ from the expected scenario, which is not as well suited for our setting, in which the model uses forecasts of future demand.
This also leads them to a different method for optimizing the robust model.

The other inputs to the model are the capacity of each unit, the LOS distribution, and various parameters such as the objective function weights.
We do not consider uncertainty in the capacity of each unit, as this can generally be known with certainty, although there can be some changes in practice such as reduced capacity caused by staff call-outs.
Uncertainty in patient LOS is captured probabilistically in the model by use of LOS distribution $\mathcal{L}$. However, there may be some uncertainty in $\mathcal{L}$ itself, particularly at the beginning of a new type of surge. Also, for small patient populations, the expected discharges may differ from true discharges significantly. However, in practice, it is a minor source of uncertainty relative to demand predictions.
Finally, there can be considerable uncertainty in the parameters, such as objective function weights. We believe that these uncertainties are best dealt with through scenario modeling and experimentation because often there is no objectively correct value -- the best parameter values are those that result in a solution that decision-makers believe is best, as they are ultimately responsible for the outcomes.
Scenario modeling can also be a powerful tool for managing uncertainty in demand and LOS distributions. Our model lends itself well to scenario modeling due to its flexibility, parameters, and fast solve times which allow for quick interactive iteration.

\subsection{Demand Forecasting}
\label{sec:methods:forecast}

Our primary aim in forecasting demand is to accurately predict the inputs to our decision optimization models: the surge patient census for capacity allocation, or surge patient arrivals for transfers. Therefore a forecast model is needed that can predict hospitalizations for the target patient population at the facility level for each hospital in the system.
The predictions must have the desired temporal resolution and horizon -- in most cases, daily predictions for two to eight weeks ahead will be most useful for tactical planning.
The predictions must also include some estimate of uncertainty, such as confidence intervals, in order to build an uncertainty set for our robust optimization model (\cref{sec:methods:uncertainty}). Without such estimates, the robust formulation of the model cannot be used, so the recommendations will not be as robust against errors in the predictions.
In addition to these requirements, it is beneficial if the predictions can be updated frequently, and if the model can generate predictions for different possible demand scenarios that decision-makers may want to plan for.
Finally, even with our robust formulation, the accuracy of the forecast is critical to ensure that the optimized decisions are as effective in practice as possible.

To address this challenge, we have developed and implemented a method to forecast facility-level hospital demand during surges that meets the required criteria.
We employ a deep learning based approach to maximize performance and flexibility. Deep learning models achieve state-of-the-art performance in general time series forecasting tasks, without requiring specialized knowledge or structure. This ensures that our method can be adapted to other hospitals and surge events without major changes.

In particular, we utilize a Time Series Dense Encoder (TiDE) \citep{das2023}, which is among the best performing models in the literature. It uses an encoder-decoder architecture with simple multi-layer perceptron blocks to encode the sequence of past covariates, and decode these embeddings into a fixed-length sequence of predictions.
We found this model achieved the best performance in our tests, compared to numerous other machine learning models, while requiring significantly less computational power than many other recent methods. The relatively small computational requirements can aid in implementation within hospital systems.

A separate model is trained for each hospital, time horizon, and target variable.
Each model is trained to predict admissions or census using retrospective hospital demand data and relevant covariates. The model leverages multiple covariates, including measures of historical demand, weather conditions, internet search trends, temporal context, and more to inform its predictions.
At each point in time, it incorporates data from the previous 60 days and predicts demand for the next 21 days.

In order to estimate uncertainty, which is crucial for decision-making, we perform quantile regression. Quantile regression involves predicting a set of quantile values for each time step, rather than a single point prediction, and training using a pinball loss function.
While this approach produces uncertainty estimates, the prediction intervals it determines are not well-calibrated, so they may not accurately capture the uncertainty. We therefore calibrate each of our prediction intervals in a post-processing step using conformalized quantile regression \citep{romano2019}, which updates the prediction intervals based on the deviation between predictions and true values over a portion of the time-series.

In its current form, the forecast is optimized for COVID-19 predictions, so it utilizes public data related to COVID-19 such as reported incidence, variant prevalence, and search engine trends.
It also uses public forecasts from the COVID-19 Forecast Hub's ensemble model of state-level COVID-19 hospitalizations and deaths \citep{cramer2021a} as covariates, taking advantage of the predictive power of existing forecasting efforts.
While this data is specific to forecasting COVID-19 hospitalizations, the model can be easily adapted to forecast demand using any available data sources relevant to the scenario.
We evaluate the performance of this model in \cref{sec:appendix:forecast-eval} of the appendix.

\section{Results}
\label{sec:results}

In this section, we demonstrate how our methodology could be applied to a hospital system and its potential to improve capacity management during surges. We do so using retrospective data from a real hospital system during the peak of the COVID-19 pandemic. The hospital system we investigate consists of five hospitals in Maryland and Washington, DC, two of which are academic medical centers, and the remainder are community hospitals. In this analysis, we focus on the period from December 15, 2021 to February 15, 2022, a two-month period that encompasses the largest wave of COVID-19 patients that the hospital system experienced. Through a close partnership with the hospital system, we obtained accurate data for this period on COVID-19 demand and capacity at the granularity required by our models. Table \ref{tab:hospitals} provides an overview of the hospitals in the system.

\begin{table*}
	\centering
	\begin{tabular}{lll}
		\toprule
		Hospital Identifier			& Hospital Type		& Approximate Bed Count		\\
		\midrule
		H1	& Academic    & $\sim 400$   \\
		H2	& Community  & $\sim 200$   \\
		H3	& Academic    & $\sim 1000$  \\
		H4	& Community  & $\sim 200$   \\
		H5	& Community	  & $\sim 200$   \\
		\bottomrule \\
	\end{tabular}
	\caption{Overview of the hospitals in the system that we partnered with to evaluate our methodology.}
	\label{tab:hospitals}
\end{table*}

\subsection{Data and Parameters}
\label{sec:results:data}

The demand data collected from the hospital system consists of the number of patients with confirmed or suspected COVID-19 occupying an ICU bed and the number of such patients admitted to an ICU at each hospital by 5:00~pm during each day in the study period. We focus on patients requiring intensive care as it is particularly important that hospitals are adequately prepared for these patients, and ICUs are often under particularly high stress. Approximately 350 such patient admissions were identified during the two-month period of this study. The data was extracted directly from the system's Hospital Information System (HIS). The number of COVID-19 patients from this period with LOS $L \geq t$ days for $0 \leq t \leq 30, t \in \mathbb{Z}$ was also estimated using the approach described in \cref{sec:methods:data}. No patient-level data was accessed or used.

The capacity of each unit of capacity was also provided.
During the early stages of the COVID-19 pandemic, a team within the hospital system identified groups of rooms within each unit at each hospital that would be converted for COVID-19 patients at the same time due to infection controls and shared resources. We adopt these bed groups as our capacity ``units''.
These units contain 9.4 beds on average (standard deviation: 6.6).
Each of these units was also assigned a ``surge level'' -- either ``baseline'', level 1-3, or ``max'' -- which indicated the suggested order that the units should be allocated to COVID-19 patients in. The baseline level was a set of units that could be sustainably devoted to COVID-19 patients, whereas the max level consisted of all bed-groups that could be allocated to those patients without more significant changes.
From this information, we assigned priority levels to each unit and estimated appropriate relative costs of converting and using each unit.
We assume there is little cost to allocating baseline units, and these units do not count as surge capacity. There is an increasing cost for converting and using units in each subsequent surge level.
Additional parameters were set through experimentation and discussions with various stakeholders and researchers in the hospital system.

\subsection{Overview}
\label{sec:results:overview}

\begin{figure}[htbp]
	\centering
	\includegraphics[width=0.85\textwidth]{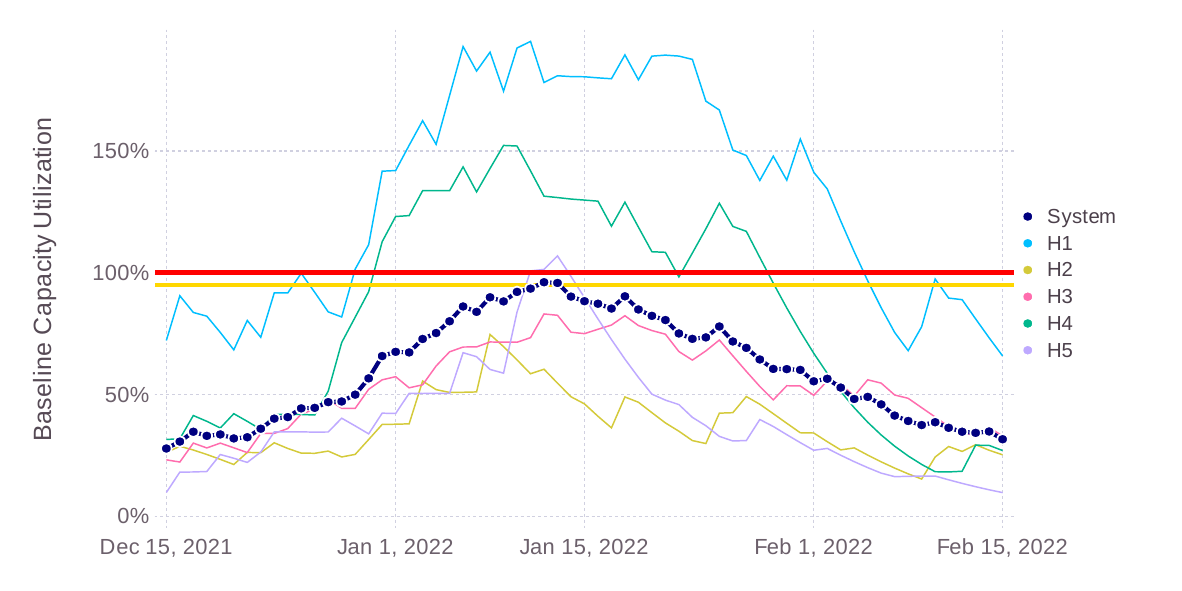}
	\caption{COVID-19 ICU occupancy of each hospital in the system relative to their baseline capacity. Values under 100\% (red line) indicate that the hospital did not require surge capacity, while values above 100\% indicate that the hospital must have activated some surge capacity. The dark blue line represents the occupancy of the system as a whole, aggregating all patients and capacity. It peaks at 97\% on January 12, 2022.}
	\label{fig:system-load}
\end{figure}

During the period we study, the COVID-19 surge had major consequences for some hospitals in the system.
Figure \ref{fig:system-load} plots the COVID-19 ICU patient occupancy relative to the baseline capacity level for the system as a whole and each hospital in it. This metric serves as a good proxy for the relative strain on each hospital caused by the influx of COVID-19 patients requiring ICU care.
This burden was extremely unequal, with H1 peaking at 195\% of baseline capacity, while H2 peaked at just 75\% of baseline.
Two of the hospitals (H2 and H3) remained well below their baseline capacities, so there was no need to consider surge capacity. Instead, the relevant decision for these facilities was whether to allocate their entire baseline capacity as a safety buffer or free up a portion for other purposes.
H5 experienced a sharp spike and briefly exceeded its baseline capacity by a small amount, making some consideration of surge capacity necessary.
Lastly, H1 and H4 went well over their baseline capacities for significant periods, requiring careful management to minimize the impacts of surge on other hospital operations, staff, and patient outcomes.
Figure \ref{fig:system-load} also reveals the system as a whole stayed barely under full utilization of baseline capacity, so if capacity was shared effectively between hospitals through transfers, little (if any) surge capacity would have been necessary.

\subsection{Dedicated Capacity Allocation}
\label{sec:results:capacity}

\begin{figure}[htbp]
	\centering
	\includegraphics[width=0.8\textwidth]{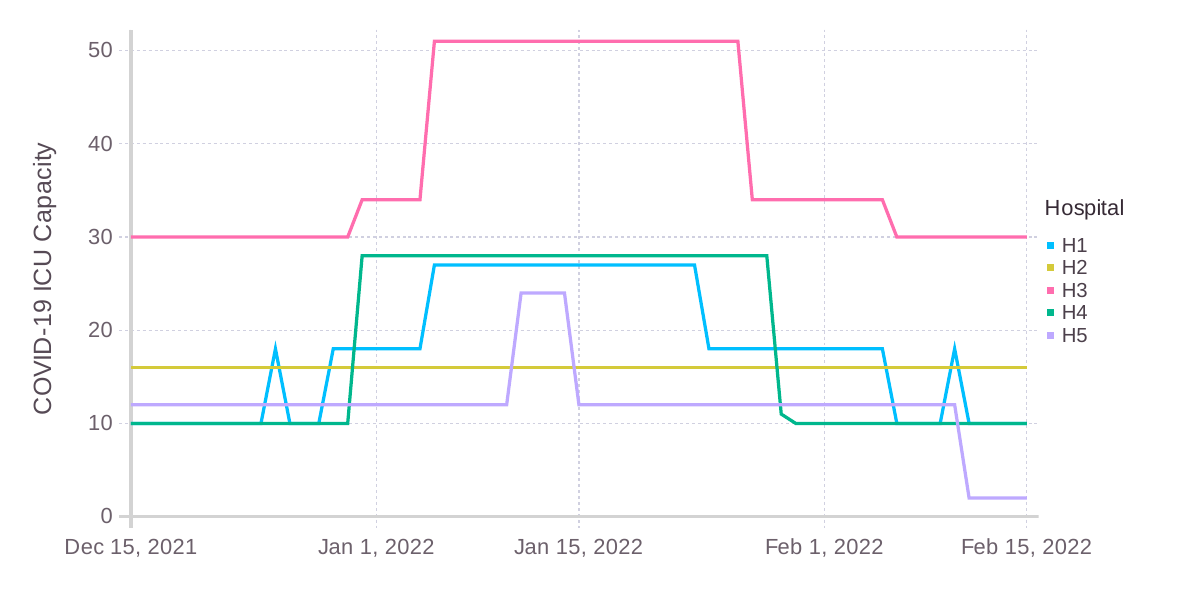}
	\caption{Amount of ICU capacity that must be allocated to COVID-19 patients at each hospital over time. Capacity cannot be adjusted by individual beds, only entire units at a time, creating discrete levels. The required capacity also reflects a maximum 95\% occupancy rate across each hospital. While H3 requires significantly more capacity allocated than the other hospitals, it can remain at baseline, as seen in \cref{fig:surge-timeline} due to also having much more capacity available.}
	\label{fig:required-capacity}
\end{figure}

Figure \ref{fig:required-capacity} plots the required capacity in each hospital over time, which is the minimum amount of capacity that the allocations must add up to. It is not simply the census of each hospital, it is computed by the optimization model, so it is based on the unit size, takes into account setup and breakdown times, and reflects that hospitals can never achieve 100\% occupancy. This is important information for hospitals because it tells them how many staff and resources they need ahead of time (if used prospectively).

\begin{figure}[htbp]
	\centering
	    \begin{subfigure}[b]{0.48\linewidth}
        \includegraphics[width=\linewidth, trim=12 15 50 30, clip=true]{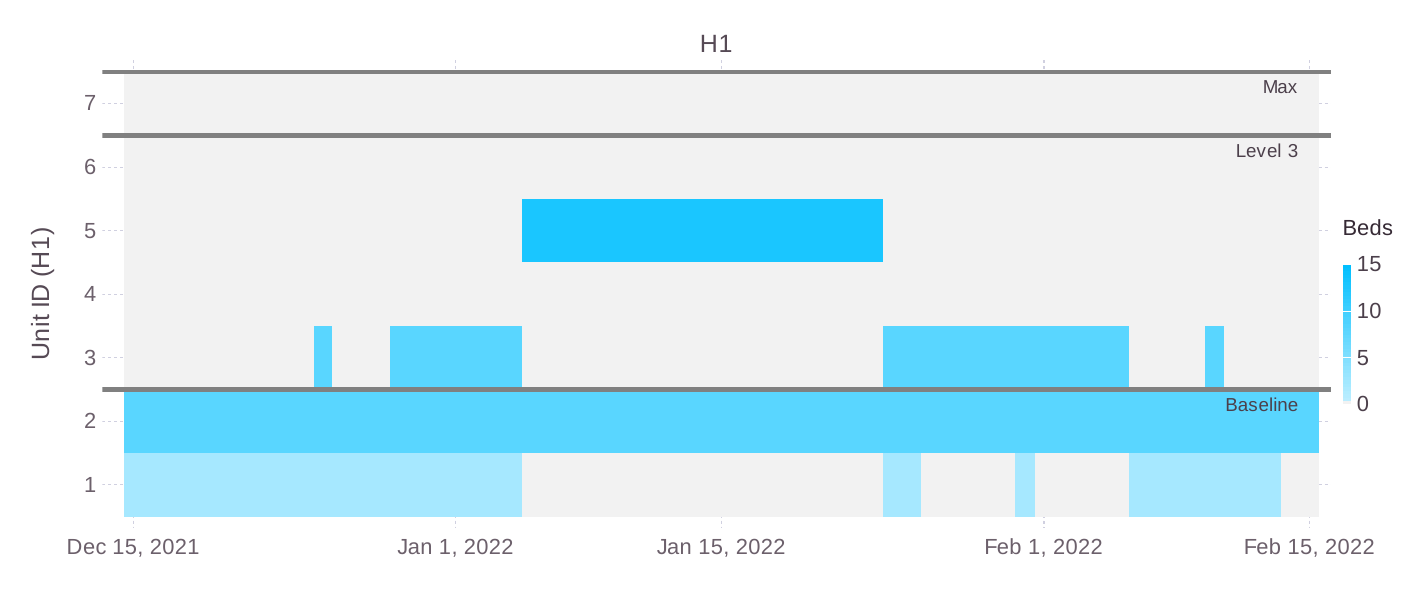}
        \caption{}
        \label{fig:unit-usage:a}
    \end{subfigure}
    \hspace{0.25em}
    \begin{subfigure}[b]{0.48\linewidth}
        \includegraphics[width=\linewidth, trim=25 15 20 30, clip=true]{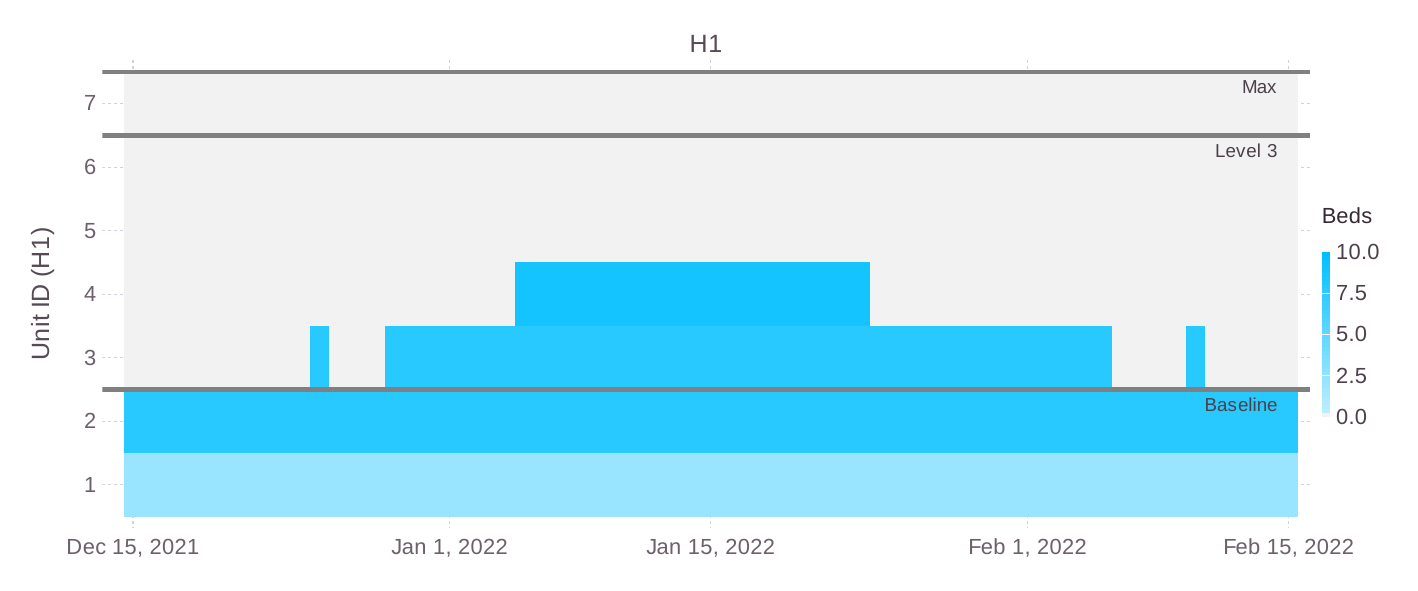}
        \caption{}
        \label{fig:unit-usage:b}
    \end{subfigure}

    \begin{subfigure}[b]{0.48\linewidth}
		\includegraphics[width=\linewidth, trim=12 15 45 30, clip=true]{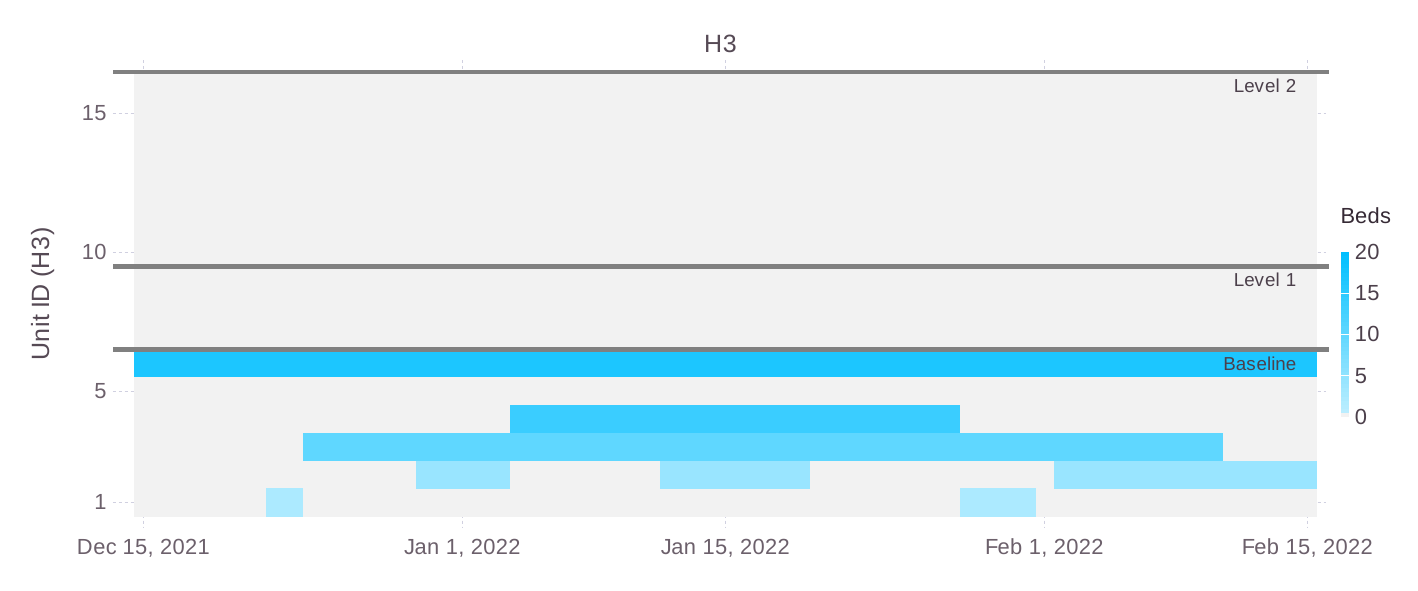}
        \caption{}
        \label{fig:unit-usage:c}
    \end{subfigure}
    \hspace{0.25em}
    \begin{subfigure}[b]{0.48\linewidth}
		\includegraphics[width=\linewidth, trim=35 15 10 30, clip=true]{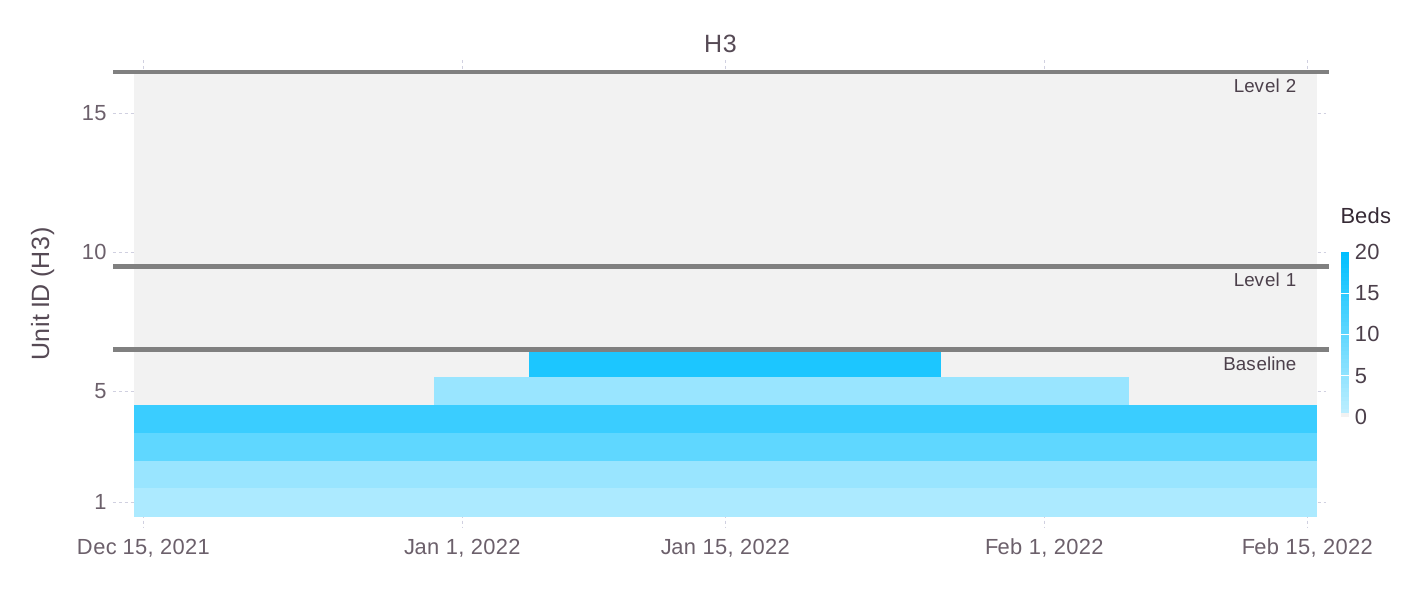}
        \caption{}
        \label{fig:unit-usage:d}
    \end{subfigure}

	\caption{Optimized ``unit'' allocations for H1 ((a) and (b)) and H3 ((c) and (d)) using no operational constraints (left) and with practical constraints (right). Each row represents a unit of capacity, and is colored blue for each period of time when it is allocated as dedicated capacity. The shade of blue indicates the number of beds in the ``unit''. Without constraints, the model sometimes chooses to use units in a higher surge level than necessary despite higher marginal costs because they fit the demand better, resulting in lower total costs. The behavior can be easily adjusted using parameters to introduce priority order, which can help improve logistics at the expense of higher objective function value.}
	\label{fig:unit-usage-base}
\end{figure}

\Cref{fig:unit-usage:a,fig:unit-usage:c} (left) now reveal the full unit allocation decisions made by the model with our default parameters. Without a constraint that units or surge levels have to be used in order, the model will sometimes choose to use an appropriately sized unit from higher surge levels rather than units from lower levels that are too big or too small for the required capacity.
\Cref{fig:unit-usage:b,fig:unit-usage:d} (right), which exhibit the optimal decisions if units have to be activated in order, observe much more stable and predictable behavior, which may benefit logistics significantly, but requires 16\% more total capacity and 11\% higher objective function value. Which of these solutions is better is ultimately up to decision-makers, but it is important to have the flexibility to meet their needs and goals.

\begin{figure}[htbp]
	\centering
	\includegraphics[width=\textwidth]{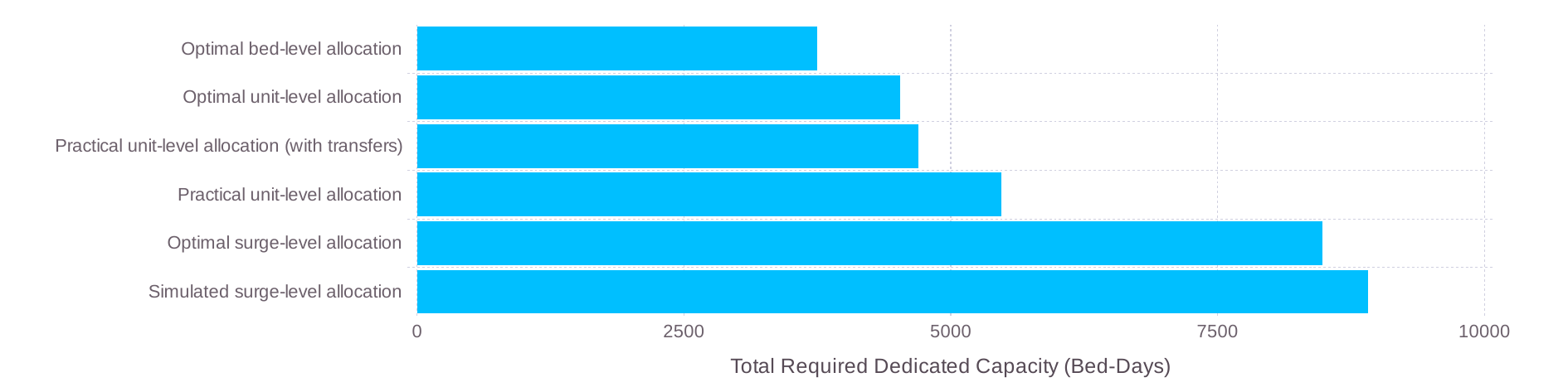}
	\caption{Total required dedicated capacity in bed-days for COVID-19 ICU patients in our case study under different allocation strategies. Optimal bed-level allocation (top) is the theoretical minimum, assuming individual beds can be allocated as needed. Our models perform optimal unit-level allocation (second bar) as well as optimal unit-level allocation that is constrained by practical considerations (third and fourth bars). We also compare against surge-level allocations (last two bars).}
	\label{fig:allocation-compare}
\end{figure}

In order to evaluate the effectiveness of these recommendations, we compare against other capacity allocation strategies in \cref{fig:allocation-compare}.
Our approach involves ``unit'' level allocations, but we also consider bed-level and surge-level allocations.
On one hand, bed-level allocations reflect the minimum required beds if dedicated capacity was not needed to ease logistical burdens, and patients could be placed immediately in any available bed. On the other hand, surge-level allocation is closer to the original planning approach of the hospital system we partnered with, which was to identify surge levels and transition between them as needed.
For surge-level allocation we consider both the optimal allocations and a simple rule-based strategy mean to represent how decision-makers might approach the problem without prospective analytics and optimization. The strategy is: if occupancy is greater than 90\%, increase the surge level so that more capacity is available if the census increases, and decrease the surge level if occupancy falls back below 70\%.

Optimized practical surge-level allocations requires approximately 40\% less dedicated capacity to serve all patients than the simulated surge-level plan, which increases to 47\% if transfers are allowed. This translates to between 54 and 66 fewer beds allocated to COVID-19 ICU patients per day on average for the hospital system. These would represent very significant improvements in surge response for the hospital system, reducing costs and stress on resources and staff.

While the minimum possible dedicated capacity using unit-level allocations is 25\% more than the minimum for bed-level allocations, it requires 46\% less than the minimum for surge-level allocations.
Furthermore, our practical unit-level allocations require only 20\% more dedicated capacity than the minimum possible for unit-level allocations. This can be further reduced using transfers, which require just 4\% more capacity than the limit, or relaxing some of the practical constraints at the cost of increased logistical challenges.

\begin{figure}[htbp]
	\centering
	\includegraphics[width=0.8\textwidth]{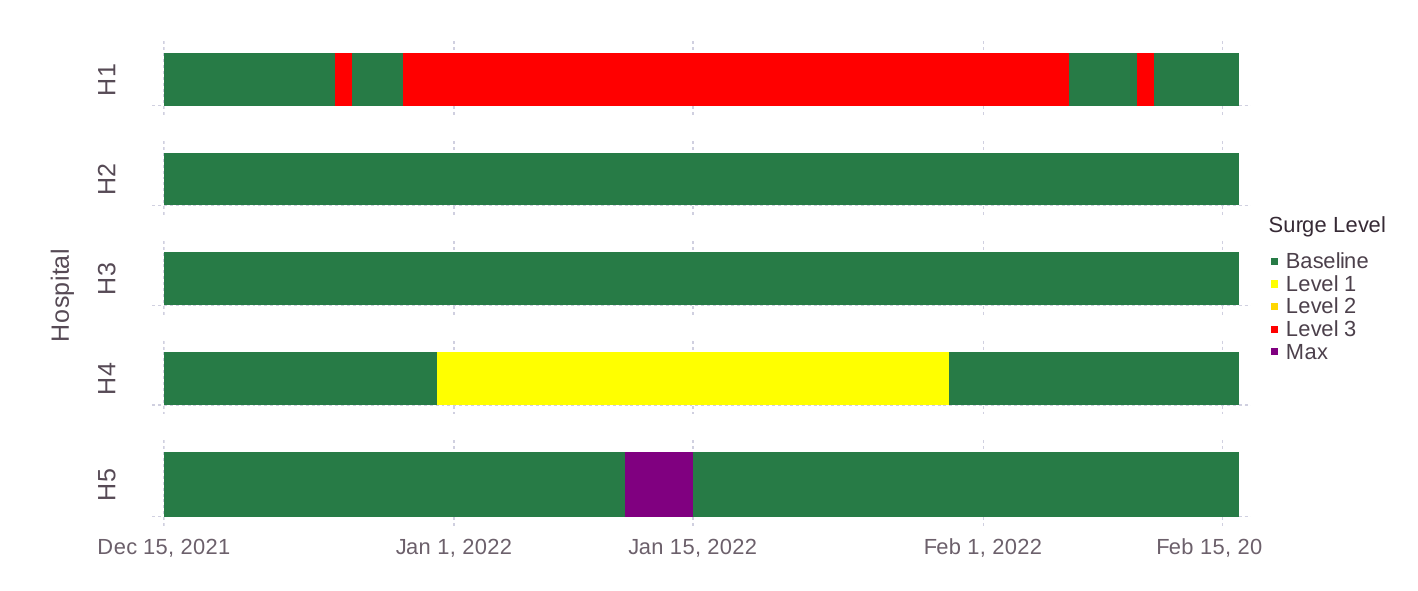}
	\caption{The timeline of maximum required surge level for each hospital, providing an overview of when each hospital needs to deploy surge capacity.}
	\label{fig:surge-timeline}
\end{figure}

Figure \ref{fig:surge-timeline} then outlines the expected surge level of each hospital over time, which is determined by the maximum surge level of the units in use. We can see that some hospitals skip directly to high surge levels because they did not have intermediate surge levels for ICUs, so they were under a lot of stress once they exceeded their baseline levels. H4 and H5 even had to use their maximum level, which underscores the need for good management of capacity utilization, and prospective analysis in particular so that they could prepare for this rapid increase in demand and provide appropriate care to all incoming patients.
Notably, H3, which is the largest and most capable hospital in the system could stay at baseline for the full surge period, meaning that its capabilities were not fully utilized even though it was the best prepared to expand into surge capacity. This highlights the potential for patient diversions from smaller, less capable hospitals to better distribute the burden caused by surges.

\subsection{Patient Transfers}
\label{sec:results:transfers}

\begin{figure}[htbp]
	\centering
	\includegraphics[width=0.8\textwidth]{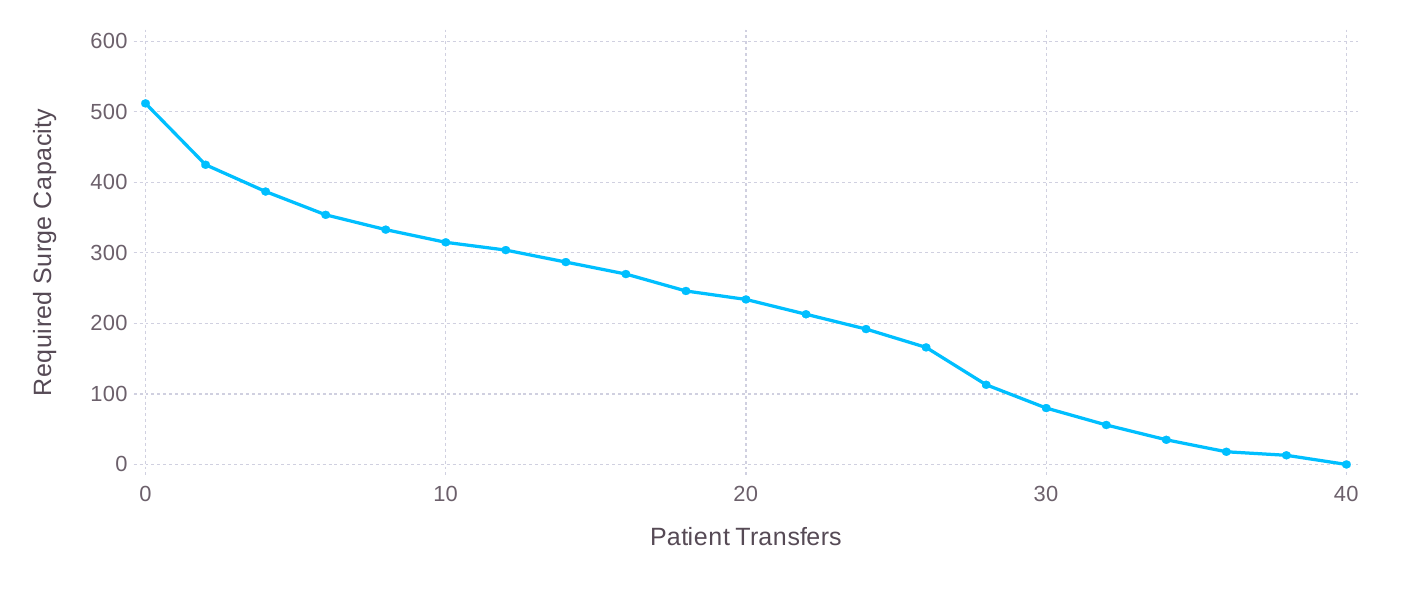}
	\caption{Number of patient transfers vs. required surge capacity (in bed-days) for the hospital system as a whole over the entire time period considered. Increasing the number of patients the model can transfer allows it to move more patients from hospitals requiring surge capacity to those that do not. Some transfers have more of an impact than others, causing some diminishing returns.}
	\label{fig:transfers-tradeoff}
\end{figure}

Transferring or diverting patients between hospitals in the system had the power to radically alter the surge requirements at each of the hospitals, depending on how many patients could have been feasibly transferred. Optimally transferring 40 patients, out of approximately 350 patient admissions ($\sim$11\%) would have been sufficient to eliminate the need for surge capacity. With 32 transfers (i.e., one transfer every two days among all five hospitals), the surge capacity requirements, measured in bed-days, can be reduced by nearly 90\% (see Figure \ref{fig:transfers-tradeoff}).

As discussed in Section \ref{sec:methods:transfers}, there is a trade-off between the number of patients transferred and the amount of surge capacity that is required. If demand across each hospital in a system is not perfectly matched to the capacities of each hospital then surge capacity may be needed at some hospitals and not others. In many cases, optimal transfers can shift incoming patients from hospitals that need surge capacity to hospitals that will not, reducing overall surge capacity usage and distributing stress more evenly. In such cases, more transfers will result in less required surge capacity, but transfers can be costly and difficult to implement at scale, so decision-makers must balance these competing objectives.
The trade-off can be made in the model using parameter $S$ in \cref{eq:cons:transfer-limits} (which limits the number of total transfers) and optimizing transfers and unit allocations to minimize required surge capacity.
Figure \ref{fig:transfers-tradeoff} illustrates the trade-off by plotting this relationship for various values of $S$.
The required surge capacity could have been reduced by 50\% using just 16 transfers if done strategically. Each transfer up reduces the surge capacity required by approximately 17 bed-days on average. However, transfers initially have an even larger impact, which slows somewhat after 10 transfers, so even a few can be highly beneficial.
The information contained in this plot is very valuable to decision-makers because it tells them with simple, clear criteria to decide how many patients to transfer: if (for example) the costs and risks of 4 transfers are less than the costs and risks of creating 400 bed-days of surge capacity then the transfers are worth doing, and otherwise, they are not.
Determining each point on the curve involves re-running the optimization model with a different number of maximum transfers, but the model is relatively efficient to solve, so it is not that time-consuming to generate enough points to infer the exact shape.

Quantifying the true utility function of capacity management decisions is impossible in practice because it requires weighing the complex financial costs of implementing decisions against the unknown potential risks to patient outcomes.
If we pose the problem as a multi-objective optimization problem, where the conflicting objectives are to minimize the costs and risks associated with creating surge capacity, and to minimize the costs and risks associated with transfers, then \cref{fig:transfers-tradeoff} represents the Pareto frontier. Each point contained in the curve is Pareto optimal, so human decision-makers must use outside information or preferences to determine the best solution out of this set in practice.
Importantly though, it provides these decision-makers with precise, data-driven information enabling them to make informed decisions.

\subsection{Implementation}
\label{sec:results:implementation}

Data for this case study was extracted from databases in the HIS used by the hospital system we partnered with using a set of SQL queries. This data was used directly by the forecasting models and optimization models. The forecasting models were implemented in Python 3.9 using Scikit-Learn \citep{pedregosa2011}. These models took approximately one minute to train for each forecast horizon. The optimization model was implemented in Julia 1.8 \citep{Julia-2017} using the JuMP modeling package \citep{lubin2022} and solved using the Gurobi MILP solver (version 9.5.2) \citep{gurobi}. The optimization model was solved to global optimality in approximately five minutes, depending on the input parameters. All figures presented in Section \ref{sec:results} were generated using the data and optimized solutions in Julia 1.8 using Gadfly \citep{jones2021}.
Code implementing our models is made available at:
\mbox{\url{https://github.com/flixpar/hospital-capacity-management-optimization}}.

\section{Discussion}
\label{sec:discussion}

\subsection*{Motivation}

The COVID-19 pandemic has exposed the need for hospitals to develop strategies to combat large demand surges. The strain created by more than six million COVID-19 patients in the US over the first three years of the pandemic has been a monumental challenge for hospitals to overcome, and has been a huge burden on healthcare systems \citep{cdc2020}.
Ongoing outbreaks of COVID-19 and other respiratory viruses make managing capacity more difficult, as hospitals must be ready for large and unpredictable outbreaks, possible resource shortages, staff getting sick, and other challenges. It also makes managing capacity more critical because a wave of hospitalizations can quickly overwhelm hospitals or entire hospital systems. It therefore is crucial that hospitals have good tools to assist them in making decisions and adapting to a variety of scenarios.

While our methodology was developed in response to the COVID-19 pandemic, it is broadly applicable in many surge settings. Due to increasing demand and constraints on capacity, stress on healthcare systems is increasing, and hospitals regularly face demand surges. During flu season or outbreaks of contagious diseases like COVID-19, hospitals can experience sudden and significant spikes in demand for medical care. Natural disasters and accidents, such as hurricanes and wildfires, can also result in injuries and illnesses that require immediate attention, and are becoming increasingly frequent. Effective contingency planning is essential for hospitals to manage surges in demand and ensure they can continue to provide high-quality care. Our model was designed to be flexible and can be easily adapted to various types of surges.

\subsection*{Implementation}

A simplified version of the optimization model described in this work was first developed as part of a practical decision-support tool for the hospital system we used as a case-study in Section \ref{sec:results}. 
This decision-support tool was developed rapidly during the early stages of the COVID-19 pandemic to help stakeholders in the hospital system make informed decisions when planning surge response strategies.
Due to the time pressures involved in developing and deploying the tool, and an emphasis on developing methods to communicate potential insights to non-technical stakeholders, the optimization model used was somewhat less realistic and informative than the model introduced in this work.
In particular, instead of optimizing unit allocations, the simplified model optimized the surge level over time, assuming that all units in a given surge level would be converted at the same time, which reduced the complexity of the problem and practical considerations.
However, this process gave us confidence that the methodology we develop in this work can produce information that has real value in practice to various stakeholders within hospital systems.

\subsection*{Limitations}

While our methodology is sound and practically valuable, it does have some limitations.
Like any prospective data-driven model, it relies on data and forecasts that may be unavailable or inaccurate. Predicting surging demand ahead of time is particularly challenging. While our robust model and probabilistic approaches mitigate some of the effects of errors in the data and forecast, they do not solve the problem completely. In particular, if the true demand deviates so much from expected that it falls outside of the predicted uncertainty range, the recommendations may not be effective. Conversely, if demand predictions are accurate, the robust optimization approach can make the recommended decisions overly conservative and therefore less useful.

There are some fundamental assumptions we make in formulating the optimization model in Section \ref{sec:methods}. While we find these assumptions to be mild and acceptable in practice, they could be limitations in extending this work to other settings.
First, we assume that all patients who arrive at a hospital can be admitted during the same day, and that all transfers can be done in a day. We also assume that only arriving patients will be transferred for operational reasons, no patients are transferred after being admitted. Finally, we assume that all patients considered in the model use the same capacity. Of course, in practice, not all patients require the same resources, for example, some patients will require ICU care. However, as long as we can partition patients requiring the same resources into separate groups, the model can be run separately on each group, as we do for COVID-19 patients requiring ICU care in Section \ref{sec:results}. Within a group of patients, some variability in resource usage is still acceptable as long as patients use the same amount of resources on average.

In addition to these modeling assumptions, our model has some restrictions. We focus on two types of decisions -- dedicated capacity allocations and patient transfers -- as they are particularly impactful, are broadly applicable across hospital systems, and have been widely used during large surges. However, there are many strategies available to hospital systems to manage capacity utilization, and decisions about implementing them were not incorporated into our optimization model, but could be in the future. Similarly, there are many possible practical considerations and constraints that hospitals could have, and we focused on the most relevant and general of these, but additional constraints could be incorporated. Finally, we limited our model to tracking objectives that could be directly estimated from input data and decisions such as required surge capacity, utilization rates, and relative costs of implementing decisions. Predicting more complex downstream impacts of decisions such as patient outcomes or actual financial costs would be valuable, but would require far more data and in-depth validation, which is outside the scope of this work.
With such a general problem and the numerous complex differences between every hospital it is impossible to capture all of the possible considerations in a single model. However, by ensuring the approach and model are flexible, our methodology can be readily adapted to new settings.

\section{Conclusion}
\label{sec:conclusion}

This study addresses the pressing need for effective capacity management tools in hospitals during periods of high demand and uncertainty. By focusing on two specific strategies, allocating dedicated surge capacity and transferring patients, the study introduces a practical, proactive, data-driven system to optimize capacity management decisions for hospital systems. The proposed methodology incorporates real-world constraints and costs associated with surge response, and ensures decisions are robust against uncertainty without being too conservative, filling gaps in existing methods. This approach can enable hospital systems to use their capacity more efficiently, ensuring that all patients receive timely and appropriate care while minimizing excess surge capacity. Ultimately, this research contributes to the development of more effective and proactive strategies for making high-stakes decisions in hospitals, helping them better cope with surges and the unpredictable nature of healthcare demand.

\newpage
\backmatter

\section*{Acknowledgments}
We would like to thank Eili Klein for his help in accessing data and the initial forecasts. 
We are also grateful to the capacity command center team at the hospital we partnered with for enabling this collaboration and for their insights. We thank the Rockefeller Foundation for their support of this work through their Covid-19 Modeling Accelerator.

\newpage
\begin{appendices}

\section{Forecast Evaluation}
\label{sec:appendix:forecast-eval}

In this section, we perform a retrospective evaluation our hospital demand forecasting method developed in \cref{sec:methods:forecast} using the case study introduced in \cref{sec:results}.
We predict COVID-19 admissions and census at each of the five hospitals in the system on each day over a 14-day horizon. Predictions are made weekly on each Monday from July 1, 2021 to May 1, 2023.
Performance is evaluated across four metrics: the weighted interval score (WIS) \citep{bracher2021}, mean absolute error (MAE), root mean squared error (RMSE), and symmetric mean absolute percent error (sMAPE).
MAE, RMSE, and sMAPE measure the performance of point predictions only, so they are computed using the predicted median. The weighted interval score is a generalization of MAE to quantile predictions, allowing us to evaluate the quality of our probablistic forecasts as well.
We report both the overall scores (\cref{tab:forecast_eval:census:summary,tab:forecast_eval:admissions:summary}) and the scores broken down by days out from prediction (\cref{tab:forecast_eval:census:steps_out,tab:forecast_eval:admissions:steps_out}).

\begin{figure*}[htpb]
    \centering
    \begin{subfigure}[b]{\linewidth}
        \includegraphics[width=\linewidth]{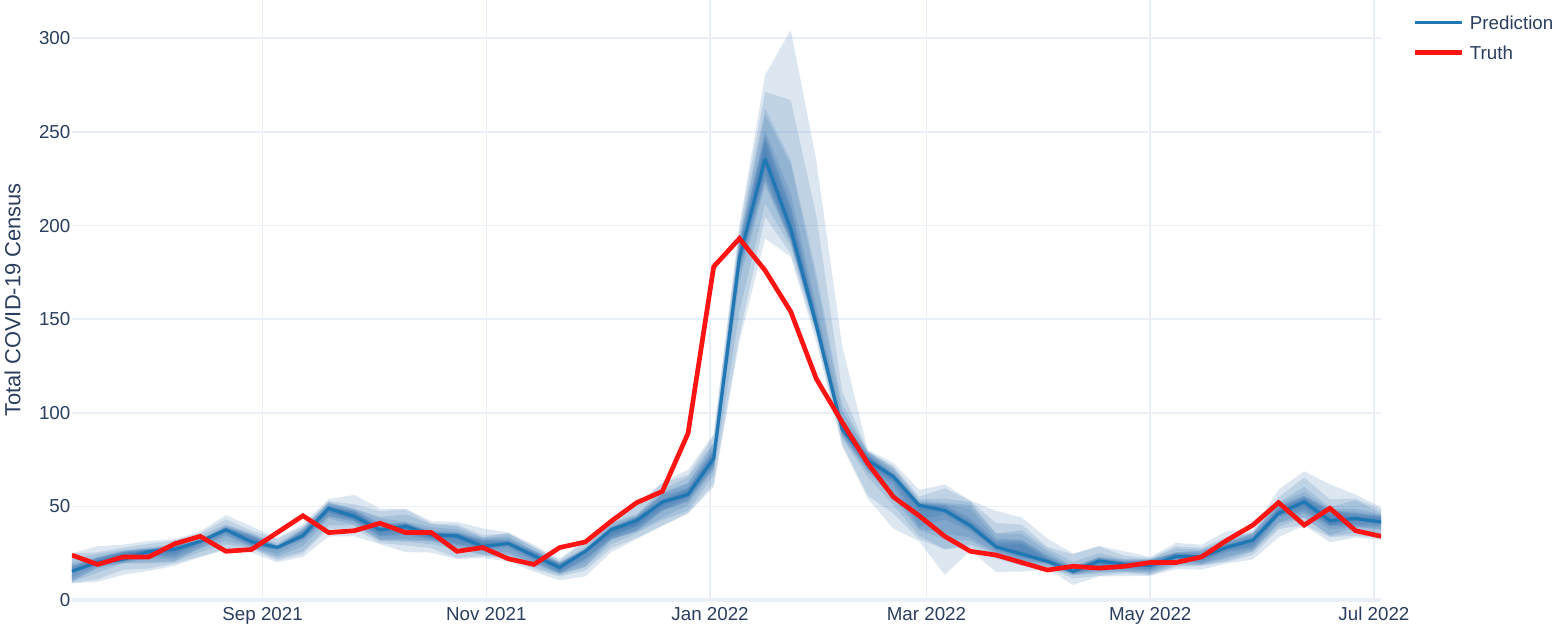}
        \label{fig:forecast:census-jhh-14daysout:surge}
        \caption{}
    \end{subfigure}
    \begin{subfigure}[b]{\linewidth}
        \includegraphics[width=\linewidth]{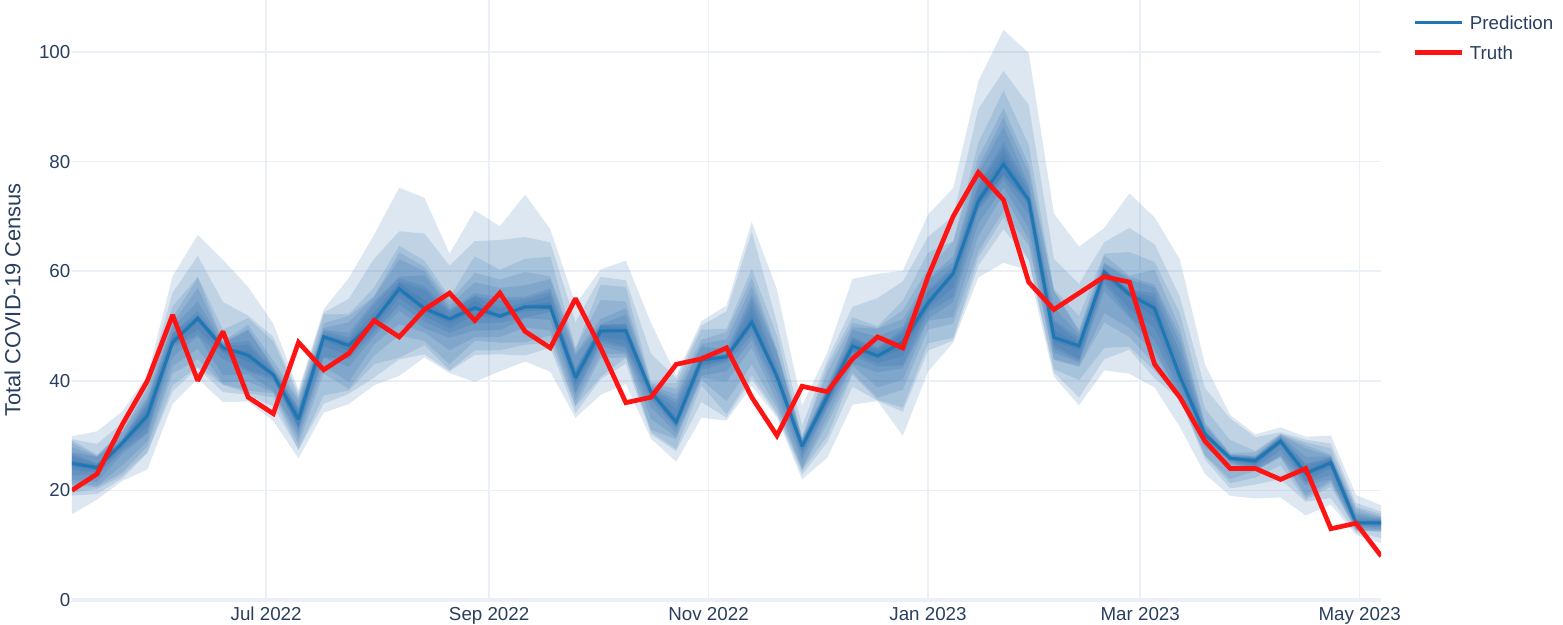}
        \label{fig:forecast:census-jhh-14daysout:nonsurge}
        \caption{}
    \end{subfigure}
    \caption{Actual (red) and predicted (blue) COVID-19 census at H3. (a) predicts from July 2021 to July 2022, whereas (b) predicts from May 2022 to May 2023. The predictions shown are 14 days out, meaning they use data up to 14 days prior to the displayed prediction. The shaded region represents the 90\% prediction interval, with darker shades representing narrower prediction intervals.}
    \label{fig:forecast:census-jhh-14daysout}
\end{figure*}

\begin{figure*}[htpb]
    \centering
    \begin{subfigure}[b]{\linewidth}
        \includegraphics[width=\linewidth]{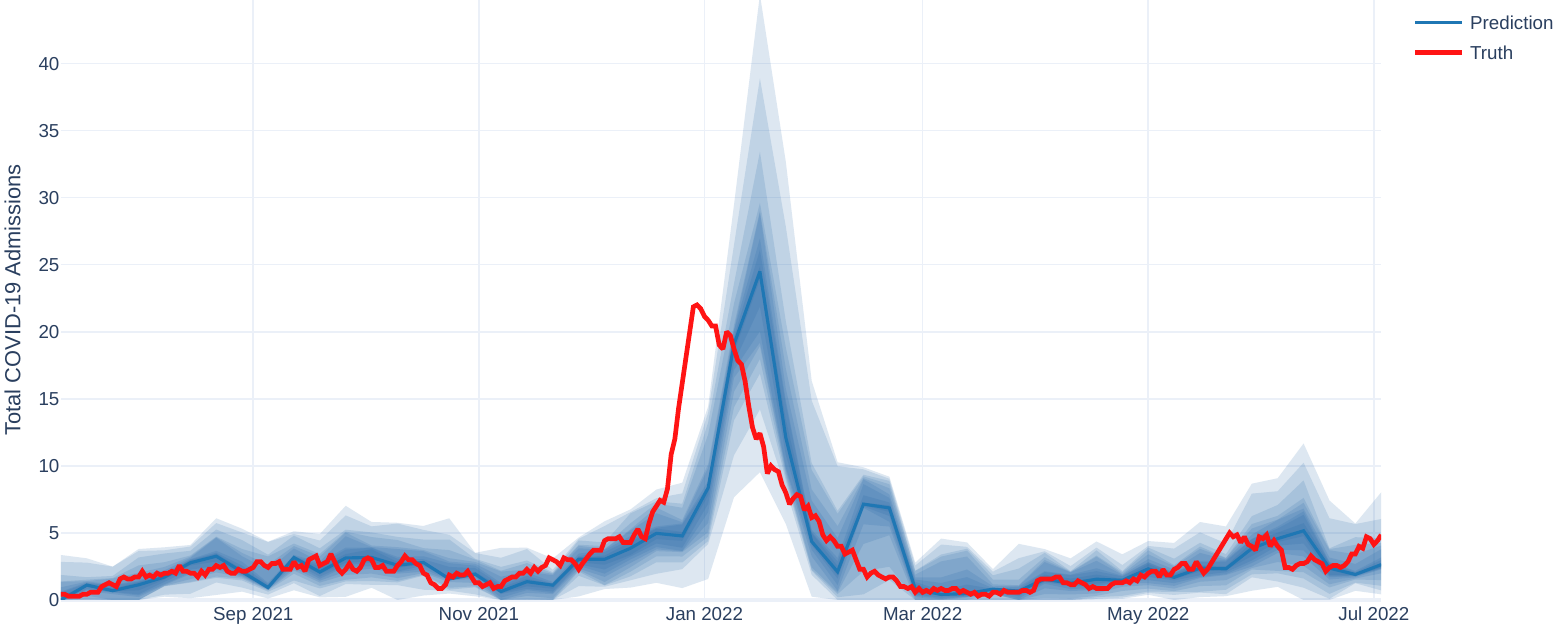}
        \label{fig:forecast:admissions-jhh-14daysout:surge}
        \caption{}
    \end{subfigure}
    \begin{subfigure}[b]{\linewidth}
        \includegraphics[width=\linewidth]{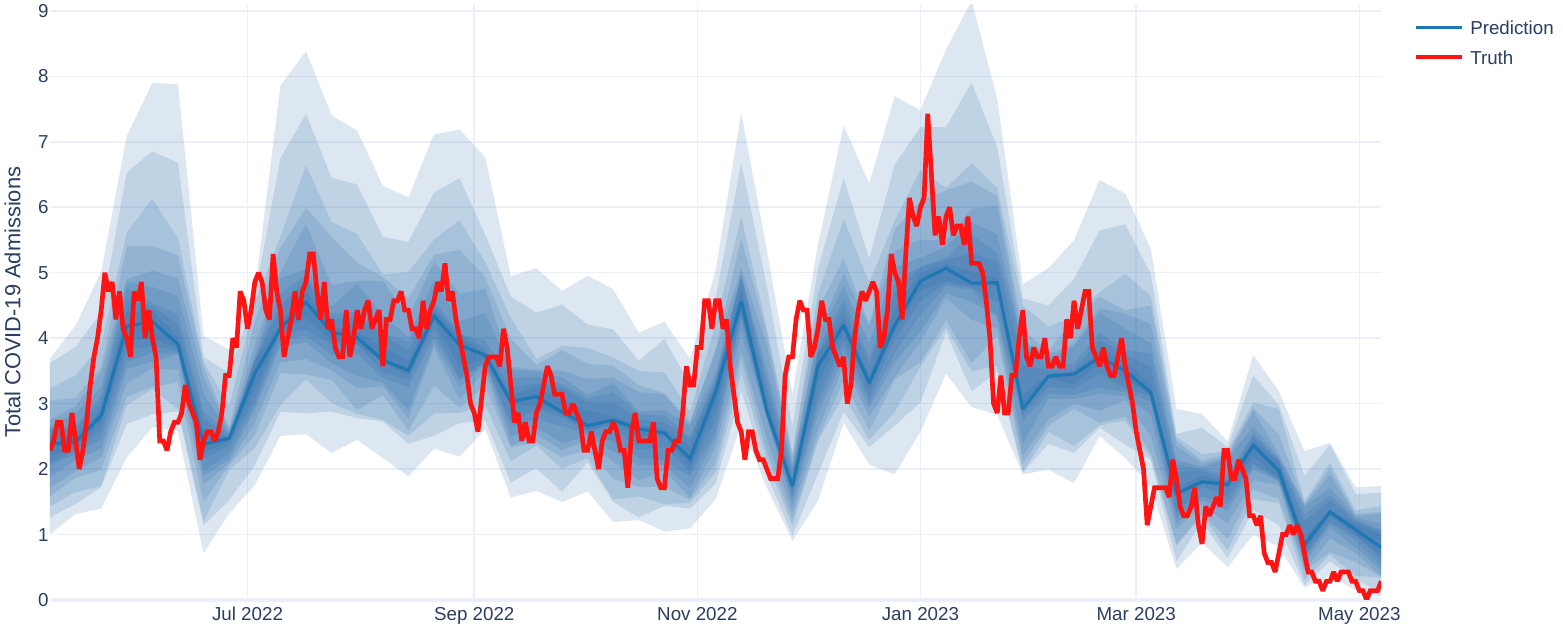}
        \label{fig:forecast:admissions-jhh-14daysout:nonsurge}
        \caption{}
    \end{subfigure}
    \caption{Actual (red) and predicted (blue) COVID-19 admissions (14 days out) at H3. \Cref{fig:forecast:admissions-jhh-14daysout:surge} plots admissions between July 2021 and July 2022, and is dominated by the largest surge of the pandemic in December 2021 through January 2022. \Cref{fig:forecast:admissions-jhh-14daysout:surge} plots admissions between May 2022 and May 2023, which contains only minor surges.}
    \label{fig:forecast:admissions-jhh-14daysout}
\end{figure*}

\Cref{fig:forecast:census-jhh-14daysout} and \cref{fig:forecast:admissions-jhh-14daysout} compare our predictions with the true values over time for COVID-19 census and admissions respectively.
Both reveal that the forecast performs very well during normal periods and the beginnings of surges, but does not keep up with the rapid explosion in hospitalizations during the worst period. This is likely because the surge period we focus on was unprecedented in scale, so the models were not trained on such extreme cases.
Nonetheless, \cref{tab:forecast_eval:admissions:summary} reports that daily COVID-19 admissions predictions were off by 1.3 patients on average, so these predictions are likely still valuable for decision-making.

\begin{table*}[htpb]
    \centering
    \begin{tabular}{l|l}
        \toprule
        Metric & Value \\
        \midrule
        WIS & 2.455 \\
        MAE & 3.770 \\
        RMSE & 5.186 \\
        MAPE & 21.39\% \\
        \botrule
    \end{tabular}
    \captionsetup{width=0.75\textwidth}
    \caption{Mean forecast performance for COVID-19 census predictions across all hospitals in our partner hospital system from July 2021 to May 2023.}
    \label{tab:forecast_eval:census:summary}
\end{table*}

\begin{table*}[htpb]
    \centering
    \begin{tabular}{l|lllll}
    \toprule
    Days out & WIS & MAE & RMSE & MAPE \\
    \midrule
	1         & 1.324 & 1.884 & 2.418 &  10.9\% \\
	2         & 1.641 & 2.474 & 3.334 &  13.0\% \\
	3         & 1.753 & 2.623 & 3.486 &  15.3\% \\
	4         & 2.075 & 3.188 & 4.051 &  18.4\% \\
	5         & 2.185 & 3.309 & 4.257 &  20.4\% \\
	6         & 2.361 & 3.583 & 4.761 &  22.5\% \\
	7         & 2.329 & 3.538 & 4.739 &  21.8\% \\
	8         & 2.594 & 4.056 & 5.483 &  23.0\% \\
	9         & 2.892 & 4.523 & 5.981 &  24.7\% \\
	10        & 2.952 & 4.627 & 6.231 &  26.0\% \\
	11        & 2.948 & 4.561 & 6.143 &  24.9\% \\
	12        & 3.046 & 4.681 & 6.285 &  26.3\% \\
	13        & 3.157 & 4.887 & 6.585 &  26.2\% \\
	14        & 3.114 & 4.852 & 6.515 &  26.0\% \\
    \botrule
    \end{tabular}
    \captionsetup{width=0.75\textwidth}
    \caption{Mean forecast performance for COVID-19 census predictions by days out from prediction across all hospitals in our partner hospital system from July 2021 to May 2023.}
    \label{tab:forecast_eval:census:steps_out}
\end{table*}

\begin{table*}[htpb]
    \centering
    \begin{tabular}{l|l}
        \toprule
        Metric & Value \\
        \midrule
        WIS & 0.933 \\
        MAE & 1.331 \\
        RMSE & 1.696 \\
        \botrule
    \end{tabular}
    \captionsetup{width=0.75\textwidth}
    \caption{Mean forecast performance for COVID-19 admissions predictions across all hospitals in our partner hospital system from July 2021 to May 2023.}
    \label{tab:forecast_eval:admissions:summary}
\end{table*}

\begin{table*}[htpb]
    \centering
    \begin{tabular}{l|rrrr}
        \toprule
        Days out & WIS & MAE & RMSE \\
        \midrule
		1   & 1.014 & 1.244 & 1.628 \\
		2   & 0.984 & 1.299 & 1.647 \\
		3   & 0.886 & 1.198 & 1.484 \\
		4   & 0.975 & 1.350 & 1.717 \\
		5   & 0.954 & 1.369 & 1.738 \\
		6   & 0.922 & 1.323 & 1.656 \\
		7   & 0.842 & 1.241 & 1.555 \\
		8   & 0.997 & 1.430 & 1.914 \\
		9   & 1.007 & 1.487 & 1.908 \\
		10  & 0.891 & 1.316 & 1.627 \\
		11  & 0.941 & 1.410 & 1.772 \\
		12  & 0.923 & 1.408 & 1.767 \\
		13  & 0.855 & 1.243 & 1.621 \\
		14  & 0.877 & 1.309 & 1.656 \\
        \bottomrule
    \end{tabular}
    \captionsetup{width=0.75\textwidth}
    \caption{Mean forecast performance for COVID-19 admissions predictions by days out from prediction across all hospitals in our partner hospital system from July 2021 to May 2023.}
    \label{tab:forecast_eval:admissions:steps_out}
\end{table*}

\end{appendices}

\clearpage
\bibliographystyle{unsrtnat}
\bibliography{hospital-capacity-management}

\end{document}